\documentclass[trackchanges,twocolumn]{aastex701}

\usepackage{tikz}
\usepackage{multirow}
\usepackage{makecell}
\usepackage{amsmath}
\usepackage{siunitx}
\usepackage{booktabs}
\usepackage{afterpage}
\usepackage{xcolor}

\usepackage{array}

\usepackage{array}
\newcolumntype{L}[1]{%
  >{\raggedright\arraybackslash\binoppenalty=9999\relpenalty=9999}p{#1}}

\newcolumntype{L}[1]{>{\raggedright\arraybackslash}p{#1}}
\newcolumntype{C}[1]{>{\centering\arraybackslash}p{#1}}
\newcolumntype{R}[1]{>{\raggedleft\arraybackslash}p{#1}}
\usetikzlibrary{arrows.meta, positioning, fit, calc, shapes.geometric}

\begin{document}

\title{Asteroids Impacting the Solar System Planets and the Moon. II: Comparison with Observational Impact Records}

\author[0009-0002-7847-8082]{Qifeng Cheng}
    \affiliation{Department of Physics, Duke University, Durham, NC 27708, USA}
    \email{qifeng.cheng@duke.edu}  
\author[0000-0002-4934-5849]{Daniel Scolnic}
      \affiliation{Department of Physics, Duke University, Durham, NC 27708, USA}
      \affiliation{Department of Electrical and Computer Engineering, Duke University, Durham, NC 27708, USA}
      \email{dan.scolnic@duke.edu}
\correspondingauthor{Qifeng Cheng,
\href{mailto:qifeng.cheng@duke.edu}{qifeng.cheng@duke.edu}}

\begin{abstract}
Planetary impact rates are difficult to infer from observations alone because impacts are rare, observational records are target-dependent, and converting observed flashes, bolides, and craters to an intrinsic impact rate depends on uncertain selection effects and conversion parameters. Building on the simulations in Paper I, we examine whether observational systematics can explain the order-of-magnitude mismatch between observed and modeled impact rates on Earth, and whether comparable discrepancies extend to other Solar System bodies. We convert 12 observed impact records for Earth, the Moon, Mars, and Jupiter to a common pre-atmospheric impact rate for $D>10$ m, propagate uncertainties through the conversion chain, and compare the resulting rates with the intrinsic rates derived in Paper~I. We find discrepancies for Earth (observation-to-model median ratio $2.2-21.0$), the Moon ($3.4-132.1$), Mars ($14.2-118.6$), and Jupiter ($2.6-47.7$), with the dominant source of the discrepancy differing by body. The simulated Jupiter-family comet (JFC) contribution substantially reduces the Earth discrepancy, suggesting that an additional dynamical source may help explain the gap. The lunar records disagree with each other by two orders of magnitude, with the discrepancy dominated by size-extrapolation uncertainties. The Mars mismatch is dominated by crater-to-impactor conversion. The Jupiter comparison depends mainly on uncertainties in the inferred number of bound impactors and in the observational completeness thresholds. Further investigation of this mismatch requires incorporating cometary modeling and tighter constraints on size-extrapolation scaling and improved observational completeness.

\end{abstract}


\section{Introduction} 

Observing asteroid or comet impactors before, during, or after impact improves our understanding of both impact rate and small-body dynamics, and informs defense strategies for human activities on Earth and beyond. Yet on Earth, the observed impact rate disagrees with the model predictions, for reasons that remain unresolved. Fireball and bolide records \citep{Brown2002,Silber2009,Brown2013} and debiased population models derived from telescopic surveys \citep{Brown2013,Harris2021,Deienno2025,NesvornyNEOMOD3} show an order of magnitude difference in decameter-scale impact rate \citep[][]{Chow2025decameter}. Such disagreement may reflect how modeled pre-impact objects and selection-biased post-entry events need to be treated explicitly with constraints. Furthermore, this mismatch on Earth may exemplify a broader Solar System-wide problem, one rooted in either population dynamics simulated in Paper I, or in observation-side biases such as conversion uncertainties and selection functions.

Identifying such mismatches is even harder beyond Earth, where impact records are obtained under a wider range of observational conditions.  Telescopic surveys detect pre-impact objects only if they are bright enough \citep{Jedicke2015_AsteroidsIV,VeresChesley2017a}, observed often enough \citep{VeresChesley2017b,Denneau2013}, and linkable into orbits \citep{Kubica2007, Chow_2026, Cheng_2026}. Fireball \citep{Brown2002}, bolide, and infrasound records \citep{Silber2009} constrain atmospheric energy deposition. Lunar \citep{Ortiz2006, Suggs2014_LunarFlashFlux, Bouley2012, liakos2024neliota} and Jovian \citep{2013Hueso, hueso2018small} flash detections measure radiative emission from hypervelocity impacts and require assumptions about luminous efficiency, impact velocity, density, and observing duty cycle. Martian seismic detections infer impact occurrence through seismic efficiency and often require confirmation through orbital imaging \citep{Posiolova2022_LargestMarsImpacts, bickel2025}. Fresh-crater surveys measure the remnant scars of impact events, but long-term crater populations also depend on resurfacing, degradation, secondary cratering \citep{Bierhaus2018Secondaries}, and chronology models \citep{Neukum2001, Hartmann2005_MartianCratering, Ivanov2001_MarsMoonCratering, Speyerer2016}. Comparing any impact rate derived from these records requires accounting for each system's selection function, the stage of the impact process it captures, and the size or energy range it observes.

Established conversion relations and the selection-bias corrections of each observation make such cross-record standardization possible. Earth fireball and bolide records provide atmospheric-entry rates and energies for meter- to decameter-scale impactors \citep[e.g.,][]{Brown2013}, with an energy-to-diameter relation calibrated by \citet{Brown2002}. Lunar flash programs such as NELIOTA constrain small impactors through optical emission and can be converted using the same energy-based relation \citep{liakos2024neliota}, whereas Lunar Reconnaissance Orbiter (LRO) temporal imaging measures fresh-crater production, which can be mapped to impactor size with crater-scaling laws \citep{Speyerer2016, HolsappleHousen2007_CraterScaling}. Jovian flashes record high-speed impacts whose energies convert to impactor size under assumed entry physics \citep{2013Hueso}. On Mars, fresh-crater surveys count dated craters directly, while InSight seismic detections convert seismic moment to crater diameter, and both can be processed with the crater-to-impactor scaling \citep{Daubar2013, zenhausern2024estimate}. Records probing sizes outside the decameter range additionally require power-law extrapolation to $D>10$ m, using slopes derived from the relevant literature \citep{Brown2002, Suggs2014_LunarFlashFlux, avdellidouTemperaturesLunarImpact2019}. Because these records include at least partial corrections for observational selection effects, the converted rates can be cautiously compared with modeled intrinsic rates to investigate the origin of the mismatch. 


In this work, we make two contributions. First, we convert 12 observational impact records, including Earth fireballs and bolides, lunar and Jovian impact flashes, lunar and Martian seismic detections and fresh-crater production, into equivalent pre-atmospheric impactor rates with uncertainty propagation. Second, we compare the intrinsic and observationally inferred rates to determine if discrepancies exist and examine whether any gap arises from source-population uncertainty or from observational selection and conversion uncertainties. 

\begin{figure*}[t]
\centering
\includegraphics[width=0.75\linewidth]{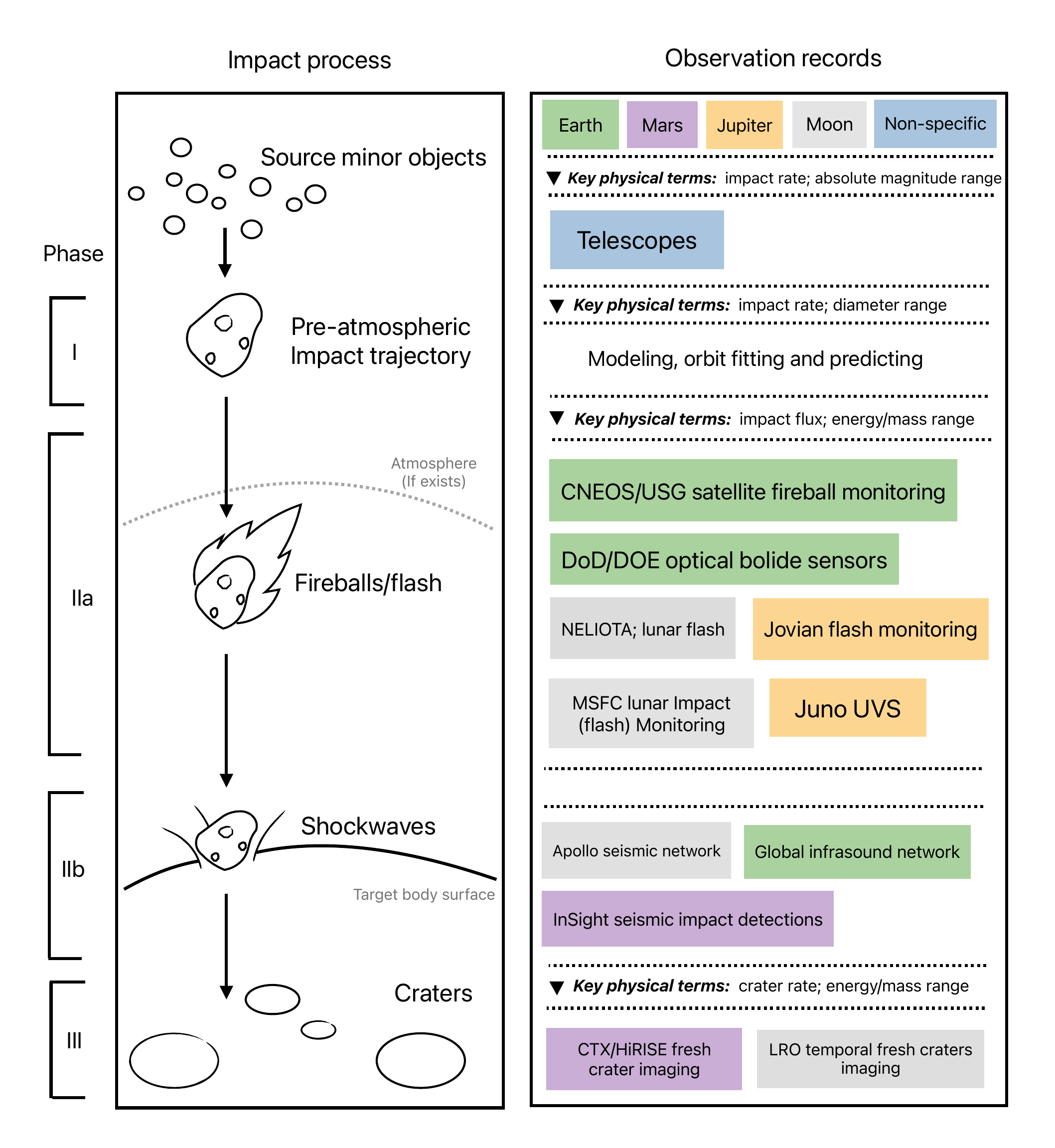}
\caption{Impact phases, records, and corresponding physical terms used in this work.}
\label{fig:observation records sum}
\end{figure*}

The paper is organized as follows. In Section~\ref{sec:method}, we explain the conversion chain, size calibration and its parameters, and uncertainty propagation for the 12 records. In Section~\ref{sec:observation comparison}, we report the observationally inferred impact rate for each record, quantify the rate discrepancy both among observational records and between observations and models, and examine the potential causes. In Section~\ref{sec:discussion}, we discuss the observation biases affecting each record, the comet-like objects that may account for the Earth rate discrepancy, and the major limitations of our work. Our main findings are summarized in Section~\ref{sec:conclusions}.

\section{Methodology}
\label{sec:method}


\begin{figure*}[t]
\centering
\includegraphics[width=0.85\linewidth]{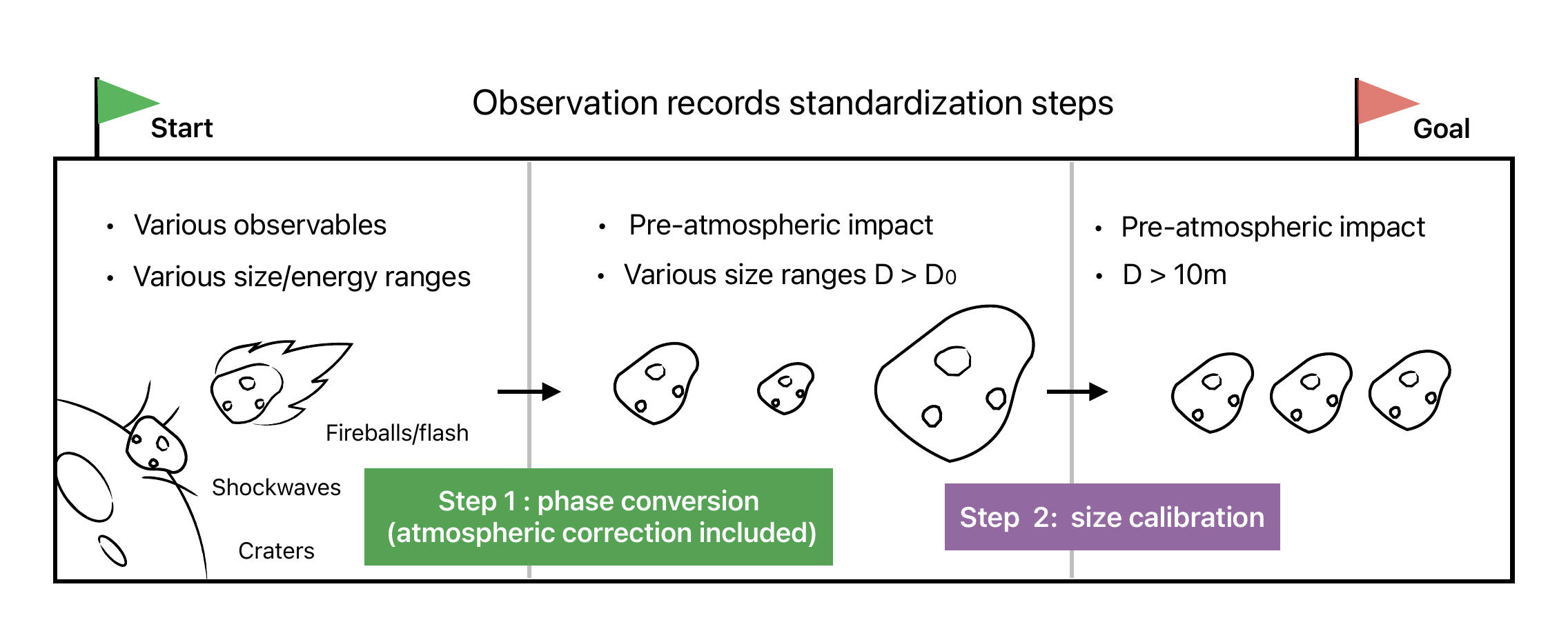}
\caption{Two-step conversion from the rates reported in the original observational records to standardized pre-atmospheric impact rates. The first step, phase conversion, converts observables measured during or after impact, such as crater formation rate or impact flash count, into the corresponding pre-atmospheric-entry impact rate. This step includes atmospheric correction where necessary; for example, the observed Martian crater-production rate must be corrected to recover the pre-atmospheric-entry impact rate. The second step, size calibration, rescales the impact rate from the size range sampled by each observational record to a common threshold of $D>10~\mathrm{m}$.}
\label{fig:standardization process}
\end{figure*}

The twelve impact records used in this work span three phases of the impact process, each involving different physical properties, as demonstrated in Figure~\ref{fig:observation records sum}. To place the various observational estimates on a common basis, we convert each to a pre-atmospheric impact rate for $D>\!10\,\mathrm{m}$, hereafter denoted $R(D>\!10\,\mathrm{m})$, through the observable conversion and size extrapolation steps illustrated in Figure~\ref{fig:standardization process}. The conversion chain therefore comprises up to three operations: (i) conversion of the observed quantity to impact rate, and of the reported impact-energy or mass range to an impactor-size range, i.e., observable to $R_0(D>D_0)$, detailed in Section~\ref{sec:method:conversion}; (ii) for post-atmospheric-entry records, an atmospheric-entry correction, detailed in Section~\ref{sec:method:atmospheric correction}; and (iii) a power-law size extrapolation to the common size threshold, i.e., $R_0(D>D_0)$ to $R(D>\!10\,\mathrm{m})$, detailed in Section~\ref{sec:method:size extrapolation}. We explain the Monte Carlo uncertainty propagation and the adopted priors in Section~\ref{sec:prior}.

\subsection{Conversion from Observable to Impact Rate}
\label{sec:method:conversion}
We convert each observational record to the same quantity: a cumulative impact rate $R$ over an equivalent diameter threshold $D_0$. This includes converting the observed quantity to an impact frequency and converting any reported energy or mass range to an equivalent diameter threshold. 

For impact records reported in terms of kinetic energy ($E_{\rm k}$), we convert energy to size using the kinetic-energy relation for a spherical impactor of bulk density $\rho_i$ and speed $v$, so that
\begin{equation}
  D = \left[12\,E_{\rm k}/(\pi\rho_i v^2)\right]^{1/3}.
\end{equation}
If the range is reported as luminous (radiated) energy $E_L$, we convert it to kinetic energy using an integral luminous efficiency $\eta$, 
\begin{equation}
  E_{\rm k} = E_{\rm L}/\eta,
  \label{eq:lumeff}
\end{equation}
with values of $\eta$ adopted from the corresponding studies. 



The observed quantity may be an impact flux or crater rate. For impact surface flux $F$, we convert to global impact rate $R$ using
\begin{equation}
  R = F \, A_{\rm p}, \qquad A_{\rm p} = 4\pi R_{\rm p}^2,
  \label{eq:wholebody}
\end{equation}
for planetary radius $R_{\rm p}$. 

For studies that report crater formation rate, namely Moon and Mars, we estimate crater-to-impactor diameter ratios using the general (combined gravity plus strength regime) scaling relations of \citet{HolsappleHousen2007_CraterScaling}, which avoids tracking the transition between the gravity- and strength-regime limits:

\begin{equation}
\label{eq:rcri_combined}
\begin{split}
\frac{D_c}{D_i}
= {} & K_1
\left[
\left(\frac{g D_i}{2 U^2}\right)
\left(\frac{\rho}{\rho_i}\right)^{2\nu/\mu}
\right. \\
& \left.
+ \left(\frac{\bar{Y}}{\rho U^2}\right)^{(2+\mu)/2}
\left(\frac{\rho}{\rho_i}\right)^{\nu(2+\mu)/\mu}
\right]^{-\mu/(2+\mu)} .
\end{split}
\end{equation}


where $D_c$ is the transient crater diameter, $D_i$ is the impactor diameter, $g$ is the surface gravitational acceleration, $\rho_i$ is the impactor density, U is the normal component of the impact velocity, $\rho$ is the target density, and $\bar{Y}$ is the effective target strength. We adopt the ``sand or cohesive soil" material constants $K_1=1.03$, $\mu=0.41$, $\nu=0.4$ for both target bodies. We adopt $\bar{Y}\sim 10\ \mathrm{kPa}$ for weak granular lunar regolith, and $\bar{Y}\sim 65\ \mathrm{kPa}$ (dry desert alluvium) for Mars \citep{williamsProductionSmallPrimary2014b}, following \citet{Holsapple1993} and \citet{HolsappleHousen2007_CraterScaling}. The impactor density and impact velocity are drawn based on the observation studies and the previous simulation results, detailed in Appendices~\ref{app:density prior} and \ref{app:velocity priors}. We apply a final-to-transient diameter ratio of $1.23-1.28$ based on the geometric collapse model \citep{GrieveGarvin1984,Johnson2016}. Because the gravity term depends on $D_i$, $D_c/D_i$ is evaluated at each record's anchor crater diameter. The anchors are: 10 m for the Moon LRO record, 3.9 m for the Mars CTX record, and 8 m for the InSight record. We report the union of the two Martian records' ranges. Our adopted parameters for each target body and resulting one-sigma {$D_c$}/{$D_i$} ranges are summarized in Table~\ref{tab:dcdi}. 


\begin{table*}
\centering
\caption{Assumed planetary parameters and crater-to-impactor scaling ranges.}
\label{tab:dcdi}
\begin{tabular}{lcccccc}
\hline
\textbf{Body} &  Record & 
$g~(\mathrm{m\,s^{-2}})$ &
$\rho~(\mathrm{kg\,m^{-3}})$&
$\bar{Y}~(\mathrm{kPa})$ & 
Anchor $D_c$ (m) & 
$D_c/D_i, 68\%$ \\
\hline
Moon & LRO & 1.62 & 1800 & $\sim10$ & 10.0 & 17.0--44.4 \\
Mars & CTX / InSight & 3.71 & 2500 & $\sim65$ & 3.9 / 8.0 & 14.8--32.1 \\
\hline
\end{tabular}
\end{table*}

These values correspond only to simple crater scaling under either gravity-dominated or strength-dominated conditions. Large complex craters may exhibit substantially larger final crater diameters due to collapse and modification processes. Venus additionally experiences strong atmospheric filtering of smaller impactors prior to surface impact. Because seismic studies typically report impact energies or crater-size ranges, we convert these observables using the same energy-to-size and crater-to-impactor relations described above. 

\subsection{Atmospheric-entry Correction}
\label{sec:method:atmospheric correction} 
The Martian records are the only ones requiring an atmospheric-entry correction, $C_{atm}$. At our adopted size range, however, the correction is negligible. Martian atmospheric deceleration and fragmentation reduce the small-crater size-frequency distribution by an order of magnitude at crater $D\sim{3}\,\rm{m}$ (impactor $D_i\sim0.2-0.3$ m), while their effect becomes negligible for $D_c>100$ m ($D_i\gtrsim 5-10$ m)
\citep{ChappelowSharpton2005,popovaBolidesPresentMartian2003,williamsProductionSmallPrimary2014b}. For the Martian records considered here,  the principal atmospheric effect on CTX/HiRISE fresh-crater measurements is fragmentation into clusters \citep{Daubar2014craters}, and the fragmentation modeling suggests that strong atmospheric filtering of the small-impactor population becomes important at crater size \(D_c\sim10\)--\(20\) cm \citep{williamsProductionSmallPrimary2014b}, outside the size range of our interest. In the Martian seismic study, \citet{Daubar2024_MarsSeismicCratering} note that atmospheric filtering can affect crater production when \(D_c\lesssim30\) m but a marked rollover is expected only below \(D_c\sim1\) m. Therefore, at our common threshold $D_i=\SI{10}{m}$, $C_{atm}\approx1$ for Mars.

\begin{figure*}[h]
\centering
\includegraphics[width=1\linewidth]{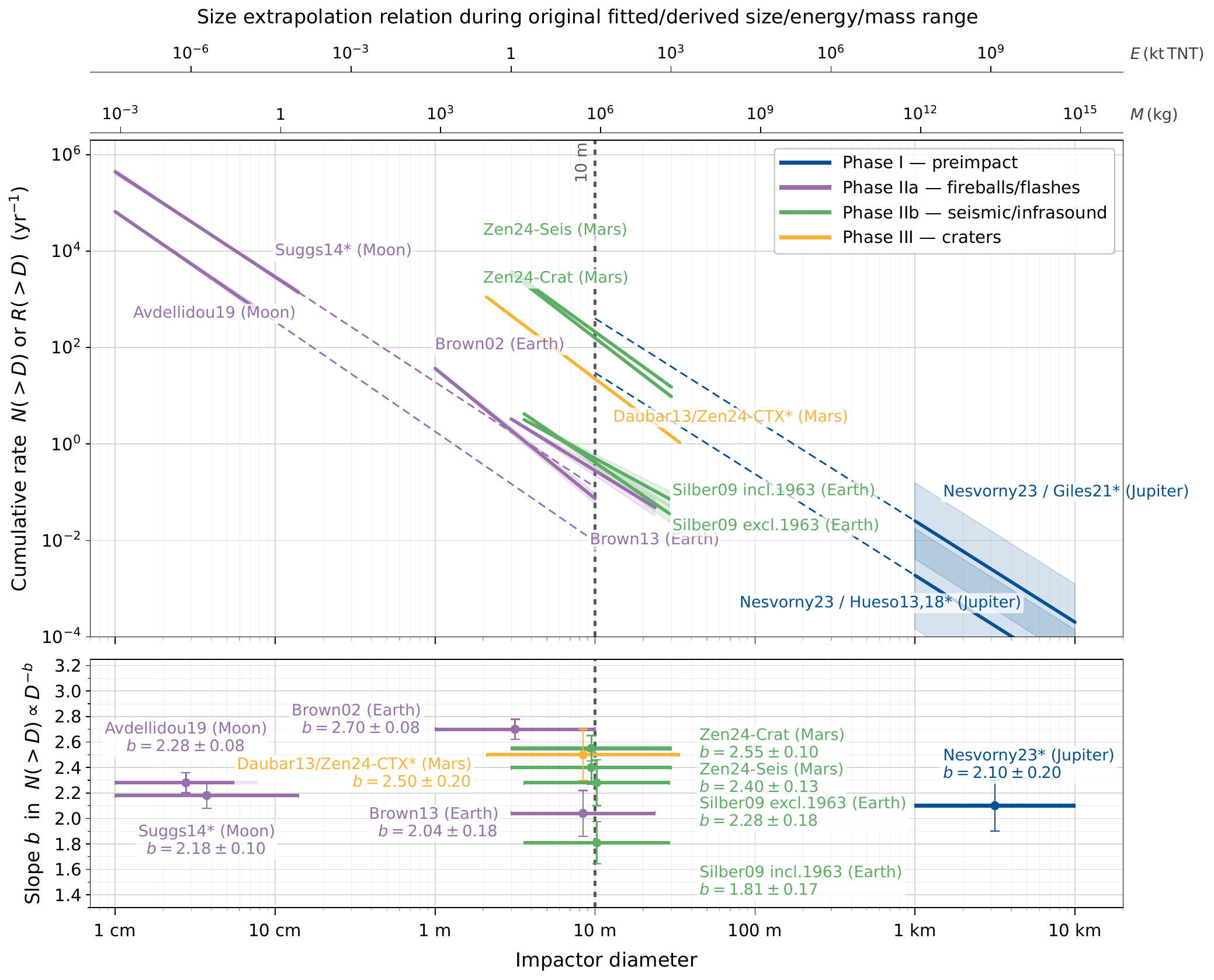}
\caption{Size-extrapolation relations adopted in this work, spanning from centimeter-scale lunar impactors to kilometer-scale Jovian ones. \textit{Top panel:} Cumulative impact rate (or frequency) $N(>D)$ or $R(>D)$ (yr$^{-1}$) as a function of equivalent impactor diameter $D$, under the assumption of a spherical impactor with bulk density $\rho = 1500$\,kg\,m$^{-3}$ and impact speed $v = 20$\,km\,s$^{-1}$. The top axes give the corresponding equivalent mass $M$\,(kg) and kinetic energy $E$\,(kt TNT).
Each solid curve spans only the size/energy/mass range over which the underlying fit or measurement was obtained. The dashed extensions represent extrapolation to this work's 10 m standardization threshold. Shaded regions indicate the uncertainty in the predicted rate: for directly fitted relations, they reflect propagated $1\sigma$ uncertainties in the slope and normalization. For Silber09 \citep{Silber2009}, ``incl.1963'' and ``excl.1963'' are the two refits of the global infrasound bolide flux with and without the single uncorroborated 3 August 1963 event; for the Mars seismic rate, ``Zen24-Seis'' and ``Zen24-Crat'' are the two independent approaches (seismic detection versus direct crater counting) of \citet{zenhausern2024estimate}, evaluated at their shared reference diameter of 8 m; for the Nesvorny23 \citep{Nesvorny2023_OuterSolarSystemImpacts} curves, they span the range of normalizations consistent with different anchor observations, such as Hueso 2013/2018 flash rates \citep{2013Hueso, hueso2018small} and Giles/Juno impact sizes \citep{giles2021detection}, with the slope fixed at $b = 2.10$.
\textit{Bottom panel:} Adopted cumulative power-law slope $b$ (where $N(>D)\propto D^{-b}$) and the diameter range over which each relation was originally fitted from the same literature sources; error bars denote $1\sigma$ uncertainties where available. An asterisk ($*$) indicates that the rate shown was not directly fitted or measured by that study but inferred or assumed. 
}
\label{fig:b-slope}
\end{figure*}

\subsection{Size Extrapolation}
\label{sec:method:size extrapolation}

After converting the observables to impact rates over their equivalent diameter ranges, we extrapolate each rate to $D>10$ m to derive the standardized impact rate $R(D>10\,\mathrm{m})$ using the power-law form of \citet{Brown2002}. This relation was originally developed for terrestrial small-NEO and fireball data and has been adopted by several studies on the Moon and Mars \citep{liakos2024neliota, Speyerer2016}. Assuming that this relation is also applicable to other planetary bodies, we apply
\begin{equation}
  R(>\!D) = R_0 \left(\frac{D}{D_0}\right)^{-b},
  \label{eq:powerlaw}
\end{equation}
throughout the 12 records, with the cumulative diameter slope $b$ calibrated to each record's specifics.

Three equivalent slope conventions appear in the source literature, and we convert all to the cumulative diameter slope $b$ using
\begin{equation}
  b \;=\; 5\alpha \;=\; 3\,b_E= 3\,b_M,
  \label{eq:slopeconv}
\end{equation}
where $\alpha$ is the differential absolute-magnitude slope ($\mathrm{d}N\propto10^{\alpha H}\,\mathrm{d}H$, with $D\propto10^{-H/5}$), $b_E$ is the cumulative energy slope ($R(>\!E)\propto E^{-b_E}$, with $E\propto D^3$ at fixed velocity and density), and $b_M$ is the cumulative mass slope ($R(>\!M)\propto M^{-b_M}$). For example, the terrestrial bolide value, $b=2.7\pm0.08$  \citep{Brown2002} corresponds to $\alpha=0.54$, $b_E=0.90$, and $b_M=0.90$. \citet{Speyerer2016} report the power-law slope $b_c$ of the crater-size distribution, which we do not adopt because it was fitted to a steep distribution of small craters outside the size range relevant to our extrapolation. Figure~\ref{fig:b-slope} shows the adopted $b$ values and the size, energy, or mass ranges over which they were originally fitted or derived, while the value applied to each observational record is listed in Table~\ref{tab:impact_rate_conversion}.


\subsection{Prior Assumptions for Uncertainty Propagation}
\label{sec:prior}

To characterize the impact rates beyond simple endpoint brackets, we derive posterior rate distributions by propagating uncertainties through the full conversion chain. Prior distributions for all input parameters are summarized in Appendix~\ref{app:conversion_priors} and reflect the assumptions adopted in the corresponding observational studies. Coefficients with published uncertainties are sampled from Gaussian distributions. For quantities without reported error bars, we introduce uncertainties in our analysis and label them as ``assigned". 

Density priors are either adopted from the observational records with assumed uncertainties or constructed from a mixture of the NEO, MBA, JFC, Centaur, and scattering-TNO source populations, weighted by the relative impact counts from Paper~I. The aim is to keep the density distributions as consistent as possible with the assumptions adopted in each observational study. Details of density prior constructions can be found in Appendix~\ref{app:density prior}. We construct the velocity priors from the Paper I velocity distributions, detailed in Appendix~\ref{app:velocity priors}. 

We make two additional prior choices. For the lunar flash impact flux, the size-extrapolation slope $b$ is drawn from an equal-weight, three-component mixture of $\{2.28\pm0.08,\,2.18\pm0.10,\,2.70\pm0.08\}$ \citep{Brown2002, Suggs2014_LunarFlashFlux, avdellidouTemperaturesLunarImpact2019}. For Jovian flashes, the native size threshold is determined from the sampled impact velocity rather than a random draw, i.e., $D_0=D_{0,\rm pub}\,(v_{\rm pub}/v_i)^{2/3}$ at fixed flash energy, where $D_{0,\rm pub}$ and $v_{\rm pub}$ are the threshold diameter and impact velocity assumed by \citet{hueso2018small}, and $v_i$ is the velocity sampled from our impact-velocity priors.

\section{Observation-based Impact Rate Results}
\label{sec:observation comparison}

\subsection{Observed Impact Rates}
\label{sec:calibrated_impact_rate}

In Table~\ref{tab:impact_rate_conversion}, we report the derived impact rate of $D>10$ m, together with the original observed rate and corresponding size/energy range, and the intermediate conversions. For the Moon, Mars, and Jupiter, the observed impact rates inferred from different observational records show notable internal inconsistencies.

\paragraph{Earth}
Earth's four records show the most internal consistency. They are all measured during the atmospheric-entry stage and we standardize the impact rates using $b$ values reported in each record. The derived Earth impact rate is $0.074^{+0.016}_{-0.013}~\mathrm{yr}^{-1}$ for the DoD/DOE optical-bolide record \citep{Brown2002}, $0.28\pm0.06~{\rm yr^{-1}}$ for USG and infrasound bolide records \citep{Brown2013}, $0.22\pm0.06~{\rm yr^{-1}}$ for the updated 14 USG fireball events \citep{Chow2025decameter}, and $0.50^{+0.15}_{-0.13}~{\rm yr^{-1}}$ for the infrasound record \citep{Silber2009}. The inferred rate of the infrasound record is based on the cumulative energy relation fit including a single, uncorroborated 3~August~1963 event. Excluding that event, the alternative fit gives a comparable rate of $\simeq0.41~{\rm yr^{-1}}$, so the infrasound estimate is not driven by this one anomalous detection. The impact rate from the DoD/DOE optical bolide differs the most from the others ($78\%$ lower than the mean of the other three) because it does not include later observations of large bolides, such as the Chelyabinsk airburst incorporated by \citet{Brown2013}. 

\begin{longrotatetable}
\begin{center}
\scriptsize
\begin{deluxetable}{lllllll}
\tabletypesize{\scriptsize}
\tablecaption{Converted impact-rate constraints\label{tab:impact_rate_conversion}}
\tablehead{
\colhead{Body} &
\colhead{\parbox[t]{3cm}{Study and observable}} &
\colhead{\parbox[t]{3.0cm}{Observed metric}} &
\colhead{\parbox[t]{3.0cm}{Observed range}} &
\colhead{\parbox[t]{3.2cm}{Converted impact rate\\(before extrapolation)}} &
\colhead{\parbox[t]{3cm}{Size extrapolation $b$}} &
\colhead{\parbox[t]{3.0cm}{$R(>10\,\mathrm{m})$\\$(\mathrm{yr}^{-1})$}}
}
\startdata
Moon &
\parbox[t]{3.0cm}{\cite{Suggs2014_LunarFlashFlux}; NASA/MSFC lunar impact flashes} &
\parbox[t]{3.0cm}{$F_0(M>30~{\rm g})=6.14\times10^{-10}~{\rm m^{-2}~yr^{-1}}$;\\ limiting flash energy $1.05\times10^{7}~{\rm J}$} &
\parbox[t]{2.8cm}{$M>30~{\rm g}$;\\ or $D_0=0.0386~{\rm m}$ for $\rho=1000~{\rm kg~m^{-3}}$} &
\parbox[t]{4cm}{$R_0(D>D_0)=2.33\times10^4~{\rm yr^{-1}}$} &
\parbox[t]{3cm}{$b=2.28\pm0.08$ \\ \citet{avdellidouTemperaturesLunarImpact2019}} &
\parbox[t]{3.0cm}{$(7.3^{+4.1}_{-2.6}) \times 10^{-2}$; \\ once every $\sim9-21$ yr} \\
&
&
&
&
&
\parbox[t]{3cm}{$b=2.18\pm0.10$\\ \citet{Suggs2014_LunarFlashFlux}} &
\parbox[t]{3.0cm}{$(1.28^{+0.95}_{-0.54})\times 10^{-1}$ ; \\ once every $\sim4.5-13.5$ yr} \\
&
&
&
&
&
\parbox[t]{3cm}{Alternative Earth-bolide slope: \\$b=2.70\pm0.08$\\ \citet{Brown2002}} &
\parbox[t]{3.0cm}{$(7.1^{+4.0}_{-2.5}) \times 10^{-3}$} \\
\noalign{\vskip 2pt}
\tableline
\noalign{\vskip 2pt}
Moon &
\parbox[t]{3.0cm}{\cite{liakos2024neliota}; NELIOTA lunar impact flashes} &
\parbox[t]{3.0cm}{145 sporadic flashes over 283.4 hr; whole-Moon frequency $7.4~{\rm hr^{-1}}$} &
\parbox[t]{3.2cm}{Projectile radii $0.5$--$3~{\rm cm}$;\\ map to $D_i=0.01$--$0.06~{\rm m}$; \\expected craters $\sim1.5$--$3.5~{\rm m}$} &
\parbox[t]{3.2cm}{$R_0\simeq6.49\times10^4~{\rm yr^{-1}}$ for $0.01<D_i<0.06~{\rm m}$} &
\parbox[t]{3cm}{$b=2.28\pm0.08$\\ \citet{avdellidouTemperaturesLunarImpact2019}; \\$R_0=K(D_{\min}^{-b}-D_{\max}^{-b})$}&
\parbox[t]{3.0cm}{$(9.5^{+7.1}_{-4.1})\times10^{-3}$; \\ once every $\sim60$--$190$ yr} \\
&
&
\parbox[t]{3.0cm}{whole-Moon frequency $12.6~{\rm hr^{-1}}$; including shower/stream flashes} &
&
\parbox[t]{3.2cm}{$R_0\simeq1.11\times10^5~{\rm yr^{-1}}$ for $0.01<D_i<0.06~{\rm m}$} &
\parbox[t]{3cm}{Same $b=2.28\pm0.08$;\\ finite bin} &
\parbox[t]{3.0cm}{$(1.6^{+1.2}_{-0.7})\times10^{-2}$;\\ once every $\sim35$--$110$ yr} \\
\noalign{\vskip 2pt}
\tableline
\noalign{\vskip 2pt}
Moon &
\parbox[t]{3.3cm}{\cite{Speyerer2016}; LRO fresh-crater production} &
\parbox[t]{3.0cm}{16 new impact craters with $D_c>10~\mathrm{m}$ over one-year equivalent annual search area $A_{\rm ann}=3.305\times10^{6}~{\rm km^2}$; scaled to $R_{\rm Moon}(D_c\ge10~{\rm m})\approx184\pm46~{\rm yr^{-1}}$} &
\parbox[t]{3.0cm}{$D_c\ge10~\mathrm{m}$;\\ $D_c/D_i=17.0-44.4$} &
\parbox[t]{3.2cm}{$R_0(D_i>0.59~{\rm m})\approx184\pm46~{\rm yr^{-1}}$ for $D_c/D_i=17.0$} &
\parbox[t]{3cm}{Impactor-phase extrapolation: $b=2.28\pm0.08$; $D_0=0.59~{\rm m}$} &
\parbox[t]{3.0cm}{$0.3\pm0.1$; \\once every $2.6$--$5.3$ yr} \\
&
&
&
&
\parbox[t]{3.2cm}{$R_0(D_i>0.23~{\rm m})\approx184\pm46~{\rm yr^{-1}}$ for $D_c/D_i=44.4$} &
\parbox[t]{3cm}{Impactor-phase extrapolation: $b=2.28\pm0.08$; $D_0=0.23~{\rm m}$} &
\parbox[t]{3.0cm}{$(3.2\pm1.3)\times10^{-2}$ ; \\ once every $22-51$ yr} \\
\noalign{\vskip 2pt}
\tableline
\noalign{\vskip 2pt}
Moon &
\parbox[t]{3.0cm}{\cite{lognonneMoonMeteoriticSeismic2009a}; Apollo seismic meteoritic impact flux} &
\parbox[t]{3.0cm}{Cumulative mass flux
$F_{\rm B}(>m)=F_0m^{-f}$, with $F_0=1.29\times10^3~{\rm yr^{-1}}$ and $f=0.90$} &
\parbox[t]{3.0cm}{Pre-impact projectile mass $m$ in kg; adopting $\rho_i=3000~{\rm kg~m^{-3}}$ gives $m_{10}=1.57\times10^6~{\rm kg}$} &
\nodata&
\parbox[t]{3cm}{$b_M =f=0.90$; \\ $b=3 b_M=2.70$} &
\parbox[t]{3.0cm}{$3.4\times10^{-3}$; \\ once every $\sim294$ yr} \\
\noalign{\vskip 2pt}
\tableline
\noalign{\vskip 2pt}
Earth &
\parbox[t]{3.0cm}{\cite{Chow2025decameter}; USG/CNEOS satellite fireballs} &
\parbox[t]{3.0cm}{$R_{\oplus}(D\gtrsim7.5~{\rm m})=0.467\pm0.125~{\rm yr^{-1}}$ from 14 decameter-sized impactors} &
\parbox[t]{3.0cm}{Pre-atmospheric diameter $D_0\simeq7.5~\mathrm{m}$} &
\parbox[t]{3.2cm}{$R_0(D>D_0)=0.467\pm0.125~{\rm yr^{-1}}$} &
\parbox[t]{3cm}{$b=2.70\pm0.08$} &
\parbox[t]{3.0cm}{$(2.2\pm0.6)\times10^{-1}$ ; \\ once every $\sim3.6-6.3$ yr} \\
\noalign{\vskip 2pt}
\tableline
\noalign{\vskip 2pt}
Earth &
\parbox[t]{3.0cm}{\cite{Brown2002}; DoD/DOE optical bolides} &
\parbox[t]{3.0cm}{$\log_{10}N=c_0-d_0\log_{10}D$, with $c_0=1.568\pm0.03$ and $d_0=2.70\pm0.08$} &
\parbox[t]{3.0cm}{Meter-to-decameter bolides; relation already expressed in pre-atmospheric diameter $D$} &
\parbox[t]{3cm}{$R_0=N(D>10~{\rm m})=0.074^{+0.016}_{-0.013}~{\rm yr^{-1}}$} &
\parbox[t]{3cm}{$b=d_0=2.70\pm0.08$; no additional threshold extrapolation} &
\parbox[t]{3.0cm}{$(7.4^{+1.6}_{-1.3}) \times10^{-2}$; \\ once every $\sim11.1-16.4$ yr} \\
\noalign{\vskip 2pt}
\tableline
\noalign{\vskip 2pt}
Earth &
\parbox[t]{3.2cm}{\cite{Silber2009}; \\global infrasound bolides} &
\parbox[t]{3.0cm}{$N_\oplus(>E)=4.5E^{-0.6}~{\rm yr^{-1}}$, with $E$ in kt TNT} &
\parbox[t]{3.0cm}{Adopting $v=20~\mathrm{km~s^{-1}}$ and $\rho=1500~{\rm kg~m^{-3}}$, a $D_i=10~{\rm m}$ body has $E_{10}\simeq37.5~\mathrm{kt}$} &
\parbox[t]{3.2cm}{$R_0=N(>E_{10})=0.50^{+0.15}_{-0.13}~{\rm yr^{-1}}$} &
\parbox[t]{3cm}{$b_E = 0.60$; \\$b_D=3\,b_E=1.80$;\\ no additional threshold extrapolation} &
\parbox[t]{3.0cm}{$(5.0^{+1.5}_{-1.3})\times10^{-1}$; \\ once every $\sim1.5-2.7$ yr} \\
&
&
&
&
&
\parbox[t]{3cm}{Excluding the single uncorroborated 1963 event: $N(>E)=6.5E^{-0.76}$} &
\parbox[t]{3.0cm}{$4.1\times10^{-1}$; \\ once every $\sim2.4$ yr} \\
\noalign{\vskip 2pt}
\tableline
\noalign{\vskip 2pt}
Earth &
\parbox[t]{3.0cm}{\cite{Brown2013}; USG sensor and infrasound bolides} &
\parbox[t]{3.0cm}{$N_\oplus(>E)=aE^{b_E}$, with $a=3.31\pm0.11$ and $b_E=-0.68\pm0.06$} &
\parbox[t]{3.0cm}{$E_{10}\simeq37.5~\mathrm{kt}$ for $D_i=10~{\rm m}$, $v=20~\mathrm{km~s^{-1}}$, and $\rho=1500~{\rm kg~m^{-3}}$} &
\parbox[t]{3.2cm}{$R_0=N(>E_{10})=0.28\pm0.06~{\rm yr^{-1}}$} &
\parbox[t]{3cm}{$b_E = 0.68\pm0.06$;\\ $b_D=2.04\pm0.18$} &
\parbox[t]{3.0cm}{$(2.8\pm0.6)\times10^{-1}$; \\ once every $\sim2.9-4.5$ yr } \\
\noalign{\vskip 2pt}
\tableline
\noalign{\vskip 2pt}
Mars &
\parbox[t]{3.0cm}{\cite{Daubar2013}; CTX/HiRISE fresh-crater production} &
\parbox[t]{3.0cm}{$\Phi(D_c\ge3.9~{\rm m})=1.65\times10^{-6}~{\rm km^{-2}~yr^{-1}}$} &
\parbox[t]{3cm}{$\Phi_{\rm CTX}(>D_c)\simeq1.65\times10^{-6}(D_c/3.9~{\rm m})^{-2.5}\\ \rm km^{-2}~yr^{-1}$; adopted $D_c/D_i=14.8-32.1$} &
\parbox[t]{3.2cm}{$\Phi_{\rm Mars,CTX}(D_i>10~{\rm m})\\\simeq2.7\times10^{-11}-1.7\times10^{-10}\\{\rm
  km^{-2}~yr^{-1}}$} &
\parbox[t]{3.0cm}{$b\simeq2.5$ \citep[Table~1]{zenhausern2024estimate} re-fit of \citet{Daubar2013}'s; $D_i=10~{\rm m}$ maps to $D_c=148-321~{\rm m}$} &
\parbox[t]{3.0cm}{$(0.39-2.7)\times10^{-2}$; \\ once every $\sim37-259$ yr} \\
\noalign{\vskip 2pt}
\tableline
\noalign{\vskip 2pt}
Mars &
\parbox[t]{3.0cm}{\cite{zenhausern2024estimate}; InSight seismic impacts (two independent approaches)} &
\parbox[t]{3.0cm}{Seismology: $N(D_c\ge8~{\rm m})=362\pm170~{\rm yr^{-1}}$ (their Table~1)} &
\parbox[t]{3.0cm}{$D_c\sim3$--$30~{\rm m}$; adopted $D_c/D_i=14.8-32.1$} &
\parbox[t]{3.4cm}{$R_0(D_c>8~{\rm m})=362~{\rm yr^{-1}}$} &
\parbox[t]{3cm}{$b_c=2.40\pm0.13$; $D_c/D_i=14.8-32.1$; $D_i>10~{\rm m}$ corresponds to $D_c=148-321~{\rm m}$} &
\parbox[t]{3.2cm}{$(0.51-3.3)\times10^{-1}$; \\ once every $\sim3.0-19.5$ yr} \\
&
&
\parbox[t]{3.0cm}{Cratering: $N(D_c\ge8~{\rm m})=280\pm99~{\rm yr^{-1}}$ (their Table~1)} &
&
\parbox[t]{3.4cm}{$R_0(D_c>8~{\rm m})=280~{\rm yr^{-1}}$} &
\parbox[t]{3cm}{$b_c=2.55\pm0.10$; $D_c/D_i=14.8-32.1$; $D_i>10~{\rm m}$ corresponds to $D_c=148-321~{\rm m}$} &
\parbox[t]{3.2cm}{$(0.23-1.6)\times10^{-1}$; \\ once every $\sim6-44$ yr} \\
\noalign{\vskip 2pt}
\tableline
\noalign{\vskip 2pt}
Jupiter &
\parbox[t]{3.0cm}{\cite{2013Hueso,hueso2018small}; optical impact flashes} &
\parbox[t]{3.0cm}{$R_0\simeq10$--$65~{\rm yr^{-1}}$;\\ raw discovery rate $\sim4$--$25~{\rm yr^{-1}}$} &
\parbox[t]{3.0cm}{Approximate projectile threshold $D_0=5$--$20~{\rm m}$ depending on density and velocity assumptions} &
\parbox[t]{3.2cm}{$R_0(D>D_0)=10$--$65~{\rm yr^{-1}}$} &
\parbox[t]{3cm}{JFC/ecliptic-comet slope $b=2.10$ \citet{Nesvorny2023_OuterSolarSystemImpacts}; endpoint $D_0=5~{\rm m}$} &
\parbox[t]{3.0cm}{$2.3$--$15$} \\
&
&
&
&
&
\parbox[t]{3cm}{JFC/ecliptic-comet slope $b=2.1$; endpoint $D_0=20~{\rm m}$} &
\parbox[t]{3.0cm}{$43$--$279$} \\
\noalign{\vskip 2pt}
\tableline
\noalign{\vskip 2pt}
Jupiter &
\parbox[t]{3.0cm}{\cite{giles2021detection}; Juno UVS transient flash} &
\parbox[t]{3.0cm}{One detected UV bolide; published intrinsic normalization $R_0=2.4\times10^4~{\rm yr^{-1}}$} &
\parbox[t]{3.0cm}{$m_0=250$--$5000~{\rm kg}$; \\for $\rho=250$--$2000~{\rm kg~m^{-3}}$;\\ corresponds to $D_0\simeq0.6$--$3.4~{\rm m}$} &
\parbox[t]{3.2cm}{$R_0(D>D_0)=2.4\times10^4~{\rm yr^{-1}}$} &
\parbox[t]{3.0cm}{$b=2.1$; endpoint $D_0=0.6~{\rm m}$} &
\parbox[t]{3.0cm}{$70^{+161}_{-58}$} \\
&
&
&
&
&
\parbox[t]{2.3cm}{$b=2.10$; endpoint $D_0=3.4~{\rm m}$} &
\parbox[t]{3.0cm}{$2441^{+5614}_{-2020}$} \\
\noalign{\vskip 2pt}
\enddata
\end{deluxetable}
\end{center}
\end{longrotatetable}

\paragraph{The Moon}
Lunar impact rates inferred from the two flash studies, one crater study, and one seismic study differ, and each estimate relies on extrapolation across a broad size range and is sensitive to the adopted $b$ value or crater scaling ratio. For flash detections, $b$ is the dominant factor. For MSFC lunar flash records, our derived impact rates are $0.128^{+0.095}_{-0.054}~{\rm yr^{-1}}$ and $0.073^{+0.041}_{-0.026}~{\rm yr^{-1}}$, adopting $b=2.18\pm0.10$, the original fit from \citet{Suggs2014_LunarFlashFlux}, and $b=2.28\pm0.08$, the updated fit by \cite{avdellidouTemperaturesLunarImpact2019}. Because these $b$ values were fitted at centimeter scales, far below our $D>10$ m threshold, we provide a comparison rate of $0.0071^{+0.0040}_{-0.0025}~{\rm yr^{-1}}$ using the slope of decameter Earth-bolide $b=2.70\pm0.08$ \citep{Brown2002}, given that this $b$ value was found to provide a good fit with the MSFC video-monitored impact flash observations \citep{Suggs2014_LunarFlashFlux}. The change of $b$ shifts the derived rates by two orders of magnitude. Another lunar flash record, NELIOTA \citep{liakos2024neliota}, is measured in an even smaller size regime where the fit of $b$ is poorer. Adopting $b=2.28\pm0.08$ yields an impact rate of $9.5^{+7.1}_{-4.1}\times10^{-3}~{\rm yr^{-1}}$. This extrapolation spans three orders of magnitude in diameter and changes the inferred cumulative rate by nearly seven orders of magnitude, making the result highly sensitive to the adopted value of $b$. 

The lunar crater record carries an analogous conversion sensitivity. The LRO temporal imaging detects an excess of $D_c\geq10\,\mathrm{m}$ craters \citep{Speyerer2016}, and the derived impact rate is sensitive to the crater scaling ratio $D_c/D_i$. We report impact rates of $(3.2\pm1.3)\times10^{-2}~{\rm yr^{-1}}$ for the upper-end $D_c/D_i$ of 44.4, and $ 0.3\pm0.1 ~{\rm yr^{-1}} $ for the lower-end $D_c/D_i$ of 17.0, corresponding to a factor of $\sim9$ difference in the inferred impact rate. 

By contrast, the Apollo seismic study directly derives the impactor number distribution as a function of mass from data \citep{lognonneMoonMeteoriticSeismic2009a}. Adopting this relation, we obtain an impact rate of $3.4\times10^{-3}~{\rm yr^{-1}}$. This estimate is subject to the same limitations associated with extreme upward extrapolation, because it applies the Earth-bolide energy extrapolation slope originally fitted to impactors of approximately $10^3-10^6$ kg \citep{Brown2002} to Apollo data at the $10^{-3}$ kg scale. 

Several of the lunar estimates can overlap once the extrapolation and crater-scaling uncertainties are considered. For example, the MSFC impact rate with $b=2.70$ overlaps NELIOTA's rate, the LRO impact rate at the lower end of the adopted range ($D_c/D_i=17.0$) overlaps MSFC's rate at $b=2.18$, and the LRO impact rate with $D_c/D_i=44.4$ overlaps with NELIOTA's shower-inclusive rate. However, we do not find a single choice of parameter that can reconcile all of the lunar estimates. 

Another possible source of the discrepancy among lunar impact-rate estimates is whether meteoroid-shower impactors are included in the original observations. The Apollo power-law relation was derived from data collected during non-shower periods, and NELIOTA’s rate of $7.4~\mathrm{h}^{-1}$ includes only sporadic impactors, whereas the estimates from MSFC and LRO include all detected impactors. Using NELIOTA’s total rate of $12.6~\mathrm{h}^{-1}$, which includes both sporadic and shower-associated events, increases the inferred impact rate by a factor of 1.7, but it reduces the discrepancy by only 20\%. Thus, shower inclusion contributes to the rate difference but cannot explain most of the gap, indicating that other major factors remain. The corresponding impact rate derived from the full NELIOTA sample is also reported in the table.

\begin{figure*}[t]
\centering
\includegraphics[width=0.9\linewidth]{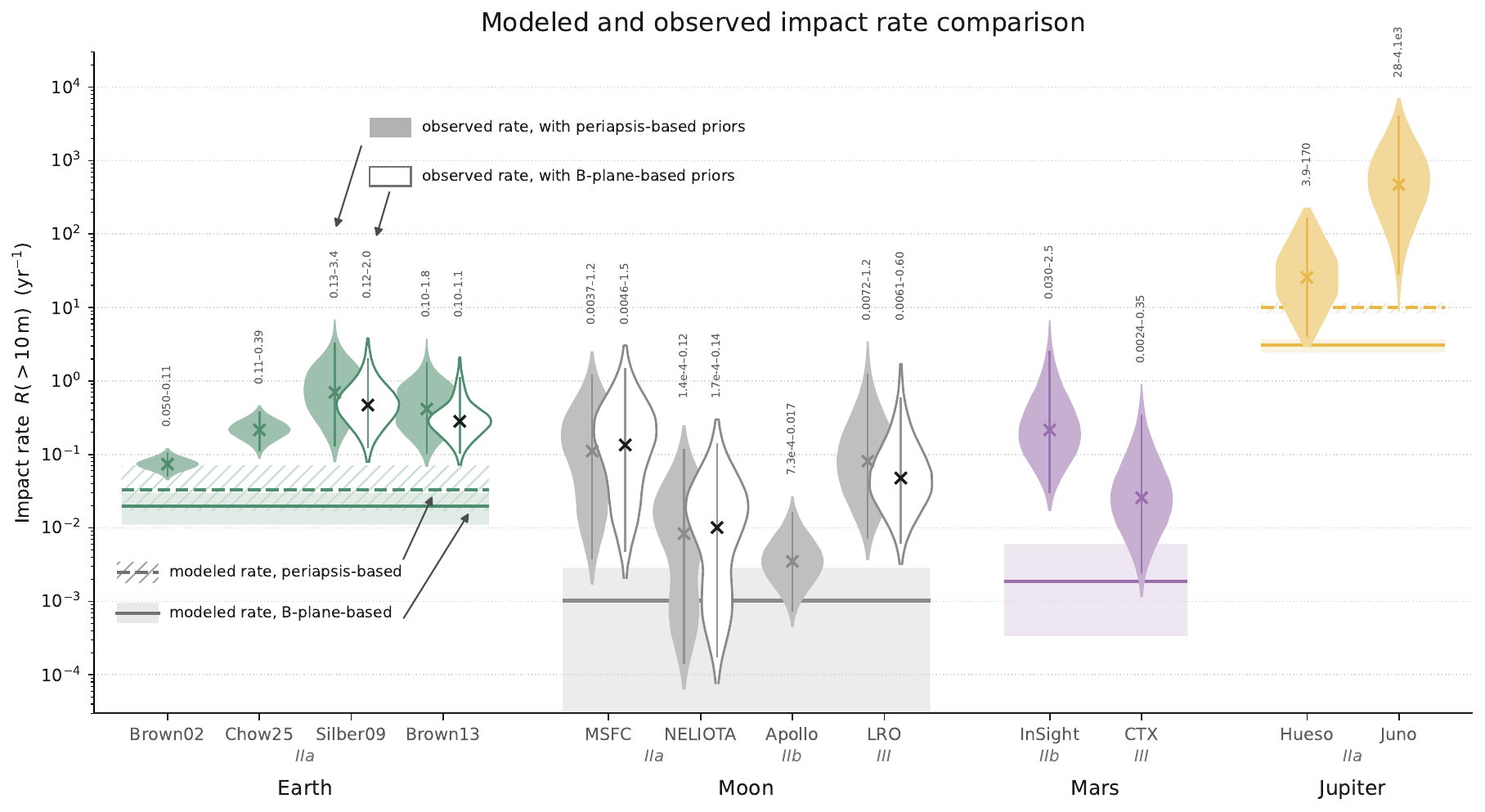}
\caption{Comparison of modeled and observed impact rates, all calibrated to the pre-atmospheric impactor threshold of $D>10$~m, for Earth, Moon, Mars, and Jupiter. Each violin shows the posterior distribution of the derived observation rate after uncertainty propagation of the full conversion chain. The cross marks the posterior median and the grey numbers give the 95\% credible interval. For records whose conversion uses velocity and density, two violins are shown, corresponding to the two choices of the simulation-derived priors. Horizontal lines show the modeled intrinsic rates derived by two methods, with shaded and hatched bands giving their $1 \sigma$ uncertainties from Paper I. The italic labels beneath the record names give the impact phase in which the original record is measured.}
\label{fig:impact_rate_all}
\end{figure*}

\paragraph{Mars}
On Mars, the derived rates similarly differ between the two observational records, and such a difference is expected. For InSight seismic data \citep{Daubar2024_MarsSeismicCratering},  our derived impact rate depends on which of their two independent approaches is used: $0.051–0.33~{\rm yr^{-1}}$ for the seismology-only approach ($b_c=2.40\pm0.13$) or $0.023–0.16~{\rm yr^{-1}}$ for the cratering-confirmed approach ($b_c=2.55\pm0.10$), each spanning the adopted $D_c/D_i=14.8–32.1$ range. For orbital imaging CTX, anchored at the best-calibrated point ($D_c=3.9$~m; \citealp{Daubar2013}), the derived rate is $0.0039-0.027~{\rm yr^{-1}}$. This is a larger gap than the reported $3-4\times$ difference between seismically derived crater rate and the traditional orbital-imaging estimation \citep{Daubar2014craters,zenhausern2024estimate}, given that we do not adopt the full least-squares re-fit cumulative size relation but anchor the relation to the original CTX data point. It reflects limitations of both records, such as the incompleteness in orbital detection of small fresh craters due to image resolution and dust coverage, and that \textit{InSight} consisted of a single seismic station with poor azimuth localization, and most detected events not confirmed as impacts \citep{zenhausern2024estimate}. We quantify how observation biases and the impact rate conversion parameters could influence the impact rate on Mars further in Section~\ref{sec:mismatch_model_obs}.

\paragraph{Jupiter}
On Jupiter, the impact rates of the two records disagree, but each has large statistical uncertainties in the original flux estimation.  For the Jupiter flashes reported by \citet{hueso2018small}, our derived impact rate spans $2.3$--$279~{\rm yr^{-1}}$, obtained by converting the endpoints of the reported $10$--$65~{\rm yr^{-1}}$ bracket for the $5$--$20~{\rm m}$ impactor range to the standardized $10$~m threshold. The original flux was estimated from three flashes, and observations were taken when Jupiter was relatively close to opposition, amateur astronomer activity was high, and geographic conditions were favorable. This results in an effective observation time of about 10-30\% of the year, clustered in Europe, Japan, North America, and Australia \citep{hueso2018small, hueso2010}. For the single detection of a bolide by Juno UVS \citep{giles2021detection}, with an estimated mass of $m = 250-5000~{\rm kg}$ (corresponding to $D \simeq 0.6-3.4~{\rm m}$ with assumed $\rho = 250-2000~{\rm kg\,m^{-3}}$),  the derived rate is $12-8055~{\rm yr^{-1}}$ after both the mass--density systematic range and the one-event Poisson uncertainty are included. Both derived rates span multiple orders of magnitude, reflecting the large uncertainty in the originally reported fluxes. 

\paragraph{Other planets}
For Mercury, Venus, Saturn, Uranus and Neptune, the existing observational studies are limited. Mercury has crater statistics from MESSENGER but cannot be converted to a present-day direct impact rate because the crater record is a long-term production and resurfacing constraint \citep{strom2011}. Venus has a well-characterized Magellan crater population, but its dense atmosphere filters small impactors and prevents crater formation below kilometer-scale crater diameters, and a 10 m projectile would likely be observed, if at all, through atmospheric entry or an airburst, neither of which has been systematically monitored \citep{Schaber1992_VenusCraters,KorycanskyZahnle2005_VenusTitan}. The giant planets beyond Jupiter lack sustained flash monitoring with the cadence and coverage needed to infer a decameter impact rate \citep{Zahnle1998_GalileanCratering,Zahnle2003_OuterSolarSystemCratering,Nesvorny2023_OuterSolarSystemImpacts,Singer2019_PlutoCharonCraters}. We therefore do not provide an observationally inferred impact rate on these bodies.

\begin{deluxetable*}{lllll}
\tablecaption{Offsets between intrinsic and observationally inferred impact rates. \label{tab:mismatch ratio}}
\tablehead{
\colhead{Target} & \colhead{$R_{\rm model}$ (yr$^{-1}$)} & \colhead{$R_{\rm obs}$ (yr$^{-1}$)} & \colhead{$R_{\rm obs}/R_{\rm model}$} & \colhead{Dominant interpretation}}
\startdata
Earth & $3.3\times10^{-2}$ & $7.4\times10^{-2}$--$7.0\times10^{-1}$ & 2.2--21.0 & \parbox[t]{2.2in}{Missing source population; updated USG record gives 6.6. }\\
Moon & $1.0\times10^{-3}$ & $3.5\times10^{-3}$--$1.4\times10^{-1}$ & 3.4--132.1 & \parbox[t]{2.2in}{Extrapolation from centimeter--decimeter projectiles and crater scaling. }\\
Mars & $1.8\times10^{-3}$ & $2.6\times10^{-2}$--$0.2$ & 14.2--118.6 & Crater/seismic conversion and completeness. \\
Jupiter & $9.9$ & $25.7$ (flashes); $471.6$ (Juno) & 2.6; 47.7 & \parbox[t]{2.2in}{Large uncertainties in observed flux estimation; Small-number event statistics.} \\[10pt]
\enddata
\tablecomments{$R_{obs}$ gives the range spanned by the posterior medians of the individual records for each body. $R_{\rm obs}/R_{\rm model}$ is derived with unrounded $R_{\rm obs}$. The full per-record 95\% credible intervals from the Monte Carlo conversion chain are shown in Figure~\ref{fig:impact_rate_all}.}
\end{deluxetable*}

\subsection{Mismatch between model and observations}
\label{sec:mismatch_model_obs}

Figure~\ref{fig:impact_rate_all} shows that the observed and modeled impact rates diverge on Earth, the Moon, Mars, and Jupiter. We plot the posterior distributions of the observed rates after propagating the conversion-chain uncertainty against the modeled rate and its uncertainty from the Paper~I simulations. Table~\ref{tab:mismatch ratio} summarizes the ratios of observed to modeled rate, $R_{\rm obs}/R_{\rm model}$, which are computed from the posterior medians, together with the candidate causes of each mismatch. Earth shows the closest agreement between model and observations, whereas Mars shows the largest discrepancy; the Moon, with four records, spans a wide range of conversion assumptions and provides the most extensively cross-checked case. The observed rate exceeds the modeled rate in every case, so we examine which uncertainties could lower $R_{\rm obs}$ or raise $R_{\rm model}$ and which could not.

\paragraph{Earth}
Earth provides the cleanest source-population testbed because fireballs and infrasound observations both detect atmospheric entries at comparable sizes. The modeled and observed impact rates disagree, with $R_{\rm obs}/R_{\rm model} \simeq 2.2-21.0$, suggesting that population-model differences may contribute to the mismatch. In Section~\ref{sec:earth mismatch}, we test whether an unmodeled cometary population could account for the gap.

\paragraph{The Moon}
Studies of the Moon provide an informative model--observation comparison because all three types of observations are available. A gap exists between models and observations with $R_{\rm obs}/R_{\rm model} \simeq 3.4-132.1$. Two records, NELIOTA flash and Apollo seismic, overlap the model predictions within their 95\% intervals, whereas the MSFC flash and LRO temporal-crater posteriors exceed the model by factors of $108.8-132.1$ and $46.7-79.3$. This excess can be a conversion artifact as discussed in Section~\ref{sec:calibrated_impact_rate}, i.e., the steeper terrestrial slope for the MSFC and NELIOTA flashes, and the larger crater-to-impactor ratio for LRO, and correcting them can bring the three rates into agreement with the modeled rate. Adding a meteor-shower population increases the observed NELIOTA rate by a factor of 1.7 (as previously calculated), which is insufficient to explain the observed-to-modeled rate mismatch.

We note that there is a difference between our simulated impact speed and the value MSFC assumes but this does not explain the model--observation gap. The simulated median lunar impact speed of $13.2~{\rm km\,s^{-1}}$ (B-plane basis) is lower than the \(24~{\rm km\,s^{-1}}\) assumed by MSFC. At fixed flash energy, this raises the inferred limiting mass by a factor of \(3.31\), the limiting diameter by a factor of $1.49$, and the extrapolated rate by approximately a factor of \(2.5\), thereby increasing the discrepancy between the MSFC rate and the model.

\paragraph{Mars}
The observed rates on Mars show the least consistency with the model, with its minimum $R_{\rm obs}/R_{\rm model}\sim14.2$, well above that of every other body ($\sim2.2-3.4$). The median CTX rate gives a mismatch factor of \(14.2\) but still overlaps the modeled rate within its uncertainty, whereas the InSight median lies a factor of \(118.6\) above the model and does not overlap it. 

Selection biases may reconcile the two records with each other but cannot reconcile the records with the model. The CTX crater record is incomplete, and a recent machine-learning search identified 123 likely new impacts and inferred a crater rate \(1.6\)--\(2.5\) times higher than earlier estimates \citep{bickel2025}. Applying this correction raises our converted CTX rate to $\simeq0.042-0.065~\mathrm{yr^{-1}}$, moving it toward the InSight value. The InSight rate, in turn, may be biased high because it assumes that the very-high-frequency event population is predominantly impact-generated \citep{zenhausern2024estimate}. Correcting either bias narrows the gap between the two records, but it shifts the observed rate further from the modeled rate. For example, applying the CTX incompleteness correction enlarges its discrepancy to a factor of roughly $23-36$. 

\(D_c/D_i\) is the parameter that most affects how well both records match the model. At the upper end of the adopted range, raising \(D_c/D_i\) from 14.8 to 32.1 lowers the cumulative CTX rate by a factor of \(\simeq6.9\) at \(b_c=2.5\) and the InSight rate by a factor of $\simeq6.4-7.2$ for its two slopes ($b_c=2.40$ and $2.55$), pushing $R_{obs}/R_{model}$ down to $\simeq2.1$ and $\simeq16.5-18.5$. Thus, selection biases can reduce the discrepancy between the two observational records, whereas uncertainty in the crater-to-impactor conversion is large enough to account for much of the discrepancy with the model. 

\paragraph{Jupiter}
For Jupiter, the inferred observation-to-model ratio is $\sim2.6$ for the Jovian impact-flash record and $\sim47.7$ for Juno/UVS, depending on the impact-rate estimator adopted from simulations. We compute these ratios using the periapsis-based direct-count estimate, which includes both bound and unbound impactors and agrees more closely with both observations than the statistically inferred rate does. This result supports the Paper~I recommendation that Jovian impact rates be interpreted primarily using direct counts.

In the observational conversion, assumptions about both impact velocity and $b$ can alter $R_{\rm obs}/R_{\rm model}$, but only the latter has a major effect. We focus here on the impact-flash record because the Juno/UVS estimate is based on a single detection. Impact velocity is important when converting between size ranges, but it is tightly constrained for Jupiter and therefore contributes little to the uncertainty in the mismatch. Both bound and unbound impactors fall within \(59.5\)--\(60.6~{\rm km\,s^{-1}}\) after accounting for the gravitational focusing of Jupiter, as discussed in Appendix~\ref{app:velocity priors}. 

The $b$ value adopted for the Jovian flash conversion, $b=2.1$, is based on a JFC-like population and may not adequately represent the NEO and MBA contributions present in the model, for which the terrestrial bolide slope ($b=2.7$) may be more appropriate. Adopting $b=2.7$ can shift the ratio, but whether it reduces or increases the discrepancy depends on the initial diameter threshold reported from the flash record \citep[$D_0 \simeq 5-20$ m;][]{hueso2018small}. At \(D_0=5~{\rm m}\), changing $b$ lowers $R_{\rm obs}/R_{\rm model}$ to $\simeq1.7$ whereas at \(D_0=20~{\rm m}\) it raises $R_{\rm obs}/R_{\rm model} $ to $\simeq4.0$. Whether the discrepancy narrows or widens depends on whether the true completeness threshold lies below or above $10~{\rm m}$. 

The dominant uncertainty in the Jovian comparison thus arises from both the modeled impact-rate estimator and the conversion of the observed flash rate. Improved constraints on flash energies and hence on impactor diameters and completeness thresholds are needed to determine whether a genuine model--observation discrepancy remains.

\section{Discussion}
\label{sec:discussion}

\label{sec:disc_conversion}

\begin{figure*}[t]
\centering
\includegraphics[width=0.85\linewidth]{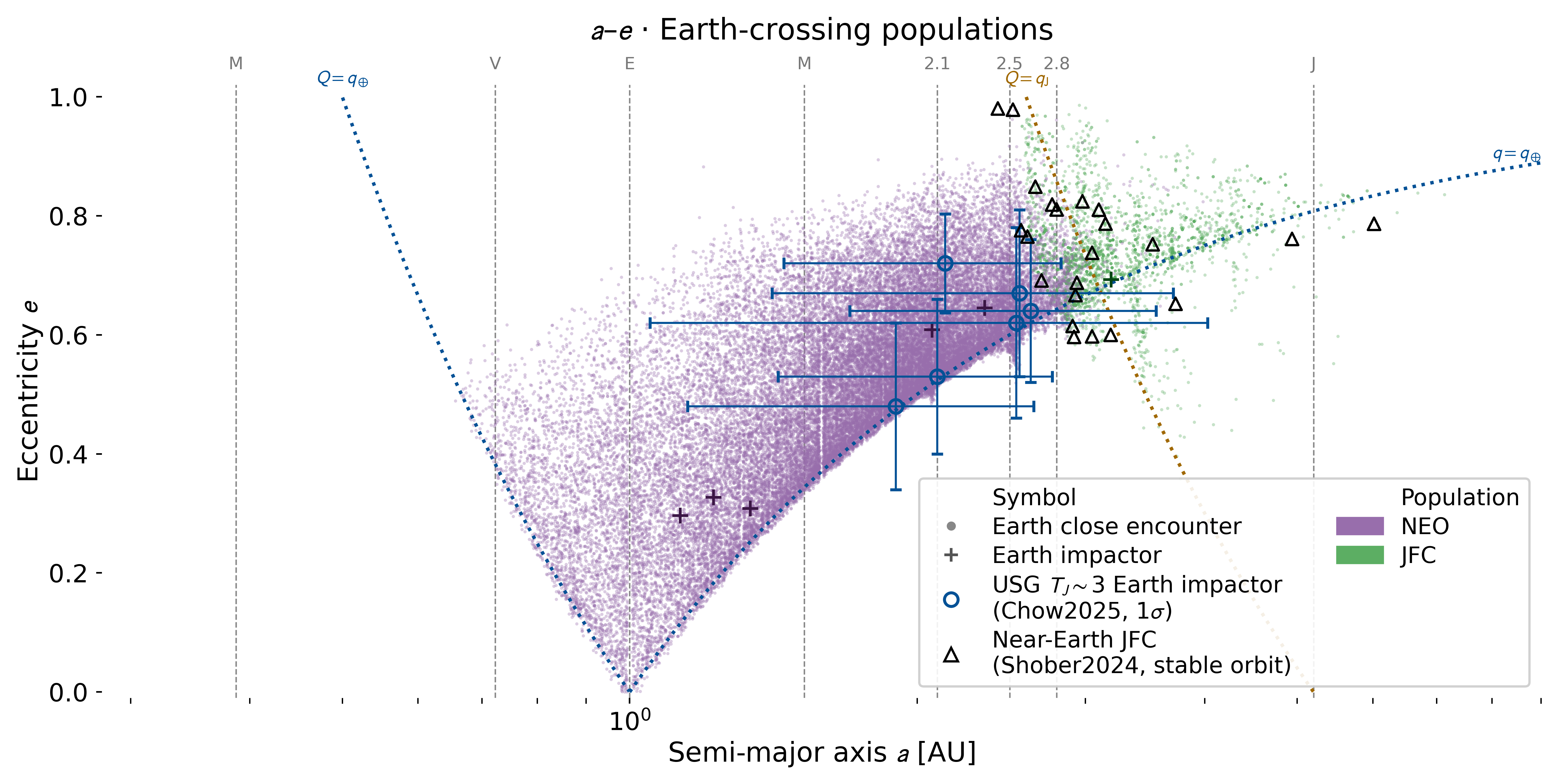}
\caption{Orbital distribution of the Earth close-encountering populations and Earth impactors from models and observations. Simulated impactors are color-coded by their source population (Purple: NEO, green: JFC); observations are taken from fireball subsets or proxies in between asteroidal and cometary origins. \label{fig:orbital jfc earth impactor}}
\end{figure*}

\subsection{Biases from Observation System Selection Function}
\label{sec:disc_selection}

Each of the 12 records has corrected some amount of observational biases. We summarize, for each record, both the correction its authors applied and the selection bias that remains. 
\paragraph{Earth} For Earth records, \citet{Brown2002} corrected the USG fireball event counts for Earth-surface coverage (60\% - 80\%) and reported a roll-off below roughly 0.1 kt as the system approaches its limiting sensitivity. Similarly for USG data, \citet{Brown2013} assumes average global coverage of about 80\% for events above 1 kt, and reports that this correction is approximate. \citet{Chow2025decameter} did not explicitly correct for coverage, but stated that the typical 7-8 kt detections are far above the 0.1 kt detection limit and thus treat it as an unbiased sample. \citet{Silber2009} corrected the historical AFTAC record for season- and hemisphere-dependent detection efficiency using an effective time--area calculation. 

\paragraph{The Moon} \citet{liakos2024neliota} corrected for effective exposure time for NELIOTA by excluding readout time and weather/technical losses, and accounted for the field of view effective image area but noted that the real coverage area is inconsistent, and the resulting flux rate is not fully debiased. \citet{lognonneMoonMeteoriticSeismic2009a} standardized the Apollo seismic record and forward modeled the instrumental and propagation response, but their analysis did not account for farside detectability, station-site-dependent effects or incompleteness at weak signals, and they identified a degeneracy between the derived impact flux and signal attenuation. \citet{Speyerer2016} corrected the LRO crater production rate for uneven area and elapsed time of each temporal image pair, and the remaining observational bias is incompleteness for small or low-contrast new craters around the imaging resolution limit ($\sim7$ m craters). \citet{Suggs2014_LunarFlashFlux} calibrated the instrumental photometry, normalized the counts by effective observing time and field-of-view area, and imposed empirical completeness limits. They did not correct for the non-uniform temporal coverage, and saturated bright flashes were reported only as lower limits.

\paragraph{Mars} \citet{zenhausern2024estimate} partly corrected the InSight seismic-impact rate by inferring the global rate from the detector's observing time and estimated detection volume. It is not fully debiased because InSight was a single-station detector, most events lack independent imaging confirmation, and the classification of very-high-frequency marsquakes as impacts remains an observational selection uncertainty. The CTX/HiRISE fresh-crater rates from \citet{Daubar2014craters,Daubar2024_MarsSeismicCratering} are normalized by repeated-image area and temporal baseline. Their remaining observational selection bias is image resolution. 

\paragraph{Jupiter} \citet{2013Hueso,hueso2018small} corrected the flash rate for observing duty cycle, projected disk coverage, and detection efficiency. Its remaining selection biases are heterogeneous amateur/professional cadence, weather, seeing, frame rate, sensitivity, longitude coverage, and event-search efficiency. The Juno UVS bolide estimate from \citet{giles2021detection} normalized coverage by using the effective UVS observing geometry and exposure. Its remaining observational selection uncertainty is dominated by the single detected event, together with UVS viewing geometry and possible non-uniform sensitivity over Jovian latitude, longitude, and observing time.

\subsection{Comet and Comet-like Earth Impactors}
\label{sec:earth mismatch}

The gap between modeled/observed impact rates on Earth is most likely to reflect a missing modeled population, and we investigate two candidate sources absent from the previous models, both are cometary-origin related. The first is the JFC-Earth impactors found by simulation in Paper~I. By adding JFC impactors, the modeled rate can increase by 50\%, closing 1/2 of the decameter gap presented by \citet{Chow2025decameter}. Although a cometary contribution to the Earth impactors has not been directly confirmed by observations, several objects from the 14 USG fireballs have large semimajor axes and eccentricities, and 6 of them have $T_J\sim3$ \citep{Chow2025decameter}, in between cometary and asteroidal origin. A comparison with the European bolide observations identified potential comet-like Earth impacts \citep{shober2024}. We show this potential alignment in the orbital parameter space in Figure~\ref{fig:orbital jfc earth impactor}, by plotting the 6 $T_J\sim3$ USG detected fireballs with the estimated $1\sigma$ uncertainties from \citet{Chow2025decameter}, the stable near-Earth JFC identified by \citet{shober2024}, and the simulated impactors and close-encountering objects from both NEOs and JFCs in Paper I. 

The second is the meteoroid-stream material such as the Taurid Complex. Steady-state NEO models such as NEOMOD3, the basis of the modeled impact rate in Paper I, include a broad cometary source statistically, but they do not explicitly preserve the angular-element and orbital-phase correlations that define individual meteor showers or resonant swarms \citep{NesvornyNEOMOD3}. Recent fragmentation and stream structure are therefore not represented explicitly, underestimating the impact rate on Earth.  Dynamical and fireball studies suggest that the 7:2 mean-motion resonance with Jupiter can concentrate Taurid material into a resonant swarm, producing recurrent close approaches to Earth and years of enhanced fireball activity \citep{Devillepoix2021}. This could introduce temporal impactor clustering that current NEO models cannot reproduce. 

The available evidence, however, does not yet establish the Taurid Complex as a sufficiently abundant decameter population to explain the remaining Earth impact-rate discrepancy. Meter-scale Taurid enhancements have been reported, but telescopic searches have not confirmed a large reservoir of tens-of-meters or larger resonant-swarm members. We plot the timeline of the known swarm-visitation periods and the fireballs and impact flashes reported, and found that none of the Earth fireball records directly fall into the roughly estimated swarm period, and that lunar records are more influenced (5 lunar flashes from NASA/MSFC and 35 lunar flashes from NELIOTA are inside the resonant swarm windows), with clustering not showing a clear trend but this is limited by the small number of reported large fireballs.

\subsection{Limitations and Future Directions}

\paragraph{Other observation records}

Several impact-monitoring records are not yet included in our work. On Earth, the Geostationary Lightning Mapper detects bolides from space with hemispheric coverage \citep{Smith2021GLM}, and dedicated ground networks such as FRIPON, the Global Fireball Observatory, and the European Fireball Network monitor entries with well-characterized exposure \citep{Colas2020FRIPON,Devillepoix2020GFO,Borovicka2022EFN}. These systems report event catalogs and small-size fluxes but do not provide a cumulative decameter rate. On the Moon, flash-to-crater linking now provides calibration of the flash impact rate conversion chain \citep{Sheward2025LIFcraters}, but the linked sample remains too small to define an independent rate. On Mars, machine-learning searches of orbital imagery continue to raise the fresh-crater detection rate \citep{bickel2025}, and this can be applied as a completeness correction to the CTX record in future steps. At Jupiter, the dedicated PONCOTS monitor detected a Tunguska-class flash, implying roughly one such event per year \citep{Arimatsu2022PONCOTS}, a constraint at energies larger than the \citet{hueso2018small} sample. A natural next step is to add these channels as new rows of Table~\ref{tab:conversion_priors} once their selection functions are published, and to forward model each observing system against the Paper~I populations. 

\paragraph{Luminous efficiency}

The luminous efficiency $\eta$ to convert flash energy to mass/size (Equation~\ref{eq:lumeff}) is a poorly calibrated parameter and the published calibrations span roughly an order of magnitude, influencing every flash-derived rate in our analysis. The Leonid-derived value of $2\times10^{-3}$ carries an order-of-magnitude uncertainty \citep{BellotRubio2000}, shower-based estimates give $(1.1$--$1.6)\times10^{-3}$ \citep{Moser2011}, crater-constrained values reach down to $7\times10^{-4}$ \citep{Madiedo2015Geminids}, and the recent flash-linked crater sample gives $\eta=(6.0\pm1.2)\times10^{-3}$ \citep{Sheward2025LIFcraters}. Our adopted prior (lognormal distribution, median $1.5\times10^{-3}$, factor of 3 at $2\sigma$; Table~\ref{tab:conversion_priors}) covers the central estimates. Because $D\propto\eta^{-1/3}$, the standardized rate scales as $R(>10~{\rm m})\propto\eta^{-b/3}$, so an order-of-magnitude $\eta$ uncertainty propagates to a factor of $\simeq6$--$8$ in the extrapolated rate for $b=2.28$--$2.70$. For example, adopting the \citet{Sheward2025LIFcraters} value, a factor of 4 above our median, would lower the MSFC and NELIOTA rates by factors of $\simeq2.9$--$3.5$ and reduce their excess over the model as in Table~\ref{tab:mismatch ratio}. The next step is to replace the uniform global $\eta$ with a record-specific value. 

\paragraph{Crater Conversion Systematics and Secondary Craters}
As listed in Appendix~\ref{app:conversion_priors}, we add arbitrary $D_c/D_i$ systematics to account for uncertain conversion assumptions, such as final-to-transient diameter ratio, the material constants $\mu,\nu,K_1$ for assumed idealized ``sand/cohesive soil" material class. After re-running the Monte Carlo process with a uniform distribution for the final-to-transient factor and $\bar Y$ over its literature range, we find neither term's contribution exceeds the assigned 20\% uncertainty. Further constraints are needed to more precisely quantify the remaining uncertainties in the crater conversion step.

Secondary craters introduce an uncertainty that our crater-to-impactor conversion cannot capture. Sub-kilometer crater populations are secondary-dominated on several bodies, with more than 95\% of Europa's small craters attributed to secondaries \citep{Bierhaus2005Europa} and a likely secondary majority below $D_c\sim1$~km on the Moon and Mars \citep{McEwenBierhaus2006,Bierhaus2018Secondaries}. Any unrecognized secondaries would overestimate the crater count, meaning that our reported observed rate may only be an upper bound.

\section{Conclusion}
\label{sec:conclusions}

We standardized twelve impact records on Earth, Moon, Mars, and Jupiter to a common pre-atmospheric impact rate at $D>10$~m, propagating the full conversion chain of energy-to-size, atmospheric, crater-scaling, and size-extrapolation priors.  The twelve records span all three impact phases, including atmospheric-entry bolides, fireballs, and infrasound on Earth \citep{Brown2002,Brown2013,Silber2009,Chow2025decameter}; impact flashes, temporal crater imaging, and seismic detections on the Moon \citep{Suggs2014_LunarFlashFlux,liakos2024neliota,Speyerer2016,lognonneMoonMeteoriticSeismic2009a}; fresh-crater imaging and seismic detections on Mars \citep{Daubar2014craters,Daubar2024_MarsSeismicCratering,zenhausern2024estimate}; and optical and ultraviolet flashes at Jupiter \citep{hueso2018small,giles2021detection}. The standardized rates (medians) are $0.074$--$0.70~{\rm yr^{-1}}$ for Earth, $3.5\times10^{-3}-0.14~{\rm yr^{-1}}$ for the Moon, $2.6\times10^{-2}$--$0.2~{\rm yr^{-1}}$ for Mars, and $25.7~{\rm yr^{-1}}$ (flashes) and $471.6~{\rm yr^{-1}}$ (Juno/UVS) for Jupiter. 

Comparison with the intrinsic rates of Paper~I reveals model--observation discrepancies on Earth, the Moon, Mars, and Jupiter. The observation-to-model ratios range from $2.2-21.0$ for Earth (6.6 for the updated USG record), $3.4-132.1$ for the Moon, $14.2-118.6$ for Mars, and $2.6-47.7$ for Jupiter. The estimates on Earth carry the smallest conversion-related uncertainties, suggesting that the discrepancy may reflect source populations that are underrepresented or absent in current models. Including a JFC component raises the modeled rate by 50\% and closes approximately half of the previously reported decameter gap. The inconsistency among the lunar impact rates lies mostly within the observations themselves, which span nearly two orders of magnitude because centimeter-scale flashes and meter-scale craters must be extrapolated over a factor of $10^{3}$ in diameter and through a crater-to-impactor ratio of $D_c/D_i=17.0-44.4$. The Martian comparison is likewise dominated by conversion uncertainties as variation in $D_c/D_i$ shifts the CTX and InSight rates both by a factor of $\simeq$7, which accounts for a moderate fraction of the model-to-observation discrepancies. At Jupiter, the direct-count intrinsic rate, including contributions from bound impactors, can reproduce the flash-derived rates when the observed initial impactor-diameter range is more tightly constrained. This agreement supports the Paper~I recommendation to favor direct counts when estimating the Jovian impact rate and highlights the need for improved constraints on observed impact energies and, consequently, impactor diameters. Tighter constraints on these conversion parameters, particularly for the Moon, Mars, and Jupiter, will be necessary to determine whether the remaining discrepancies reflect observational systematics or genuine deficiencies in the modeled impactor populations.

\begin{acknowledgments}
Q.C. and D.S. acknowledge support from the Duke University Trinity College of Arts and Sciences Department of Physics and from the Cosmology Group. D.S. acknowledges support from the Duke University Electrical and Computer Engineering Department.

D.S. is supported by the Department of Energy grant DE-SC0010007, the David and Lucile Packard Foundation, the Templeton Foundation, and Sloan Foundation.
\end{acknowledgments}

\appendix

\section{Conversion Chain Priors}
\label{app:conversion_priors}
The priors for physical quantities used in the Monte Carlo conversion chain uncertainty propagation are summarized in Table~\ref{tab:conversion_priors}. Density priors are detailed in Appendix~\ref{app:density prior} and velocity priors are detailed in Appendix~\ref{app:velocity priors}. There are a few notations not mentioned in the main text. A Gaussian prior is written as $\mathcal{N}(\mu,\sigma)$, where $\mu$ and $\sigma$ are the mean and standard deviation. A positively truncated normal distribution is written as $\rho \sim \mathcal{N}_{>0}(\mu,\sigma^2)$, where $\mu$ and $\sigma$
are the mean and standard deviation of the underlying normal distribution. A log-uniform prior between $x_{\min}$ and $x_{\max}$ has probability density $p(x)\propto x^{-1}$ over that interval. Counting uncertainty is propagated using the Jeffreys prior for a Poisson mean, $p(\lambda)\propto\lambda^{-1/2}$. For a record containing $N$ events, the resulting posterior is
\begin{equation}
\lambda\mid N\sim
\Gamma!\left(N+\tfrac{1}{2},1\right),
\end{equation}
where the second argument denotes a unit rate. When the record has exposure $T$, the sampled event rate is $\lambda/T$. When a published normalization derived from $N$ events is retained as the central value, counting uncertainty is instead introduced through the multiplicative factor $\lambda/N$. Thus, the USG rate is sampled as $\Gamma(14.5,1)/(30 \mathrm{yr})$, where the effective exposure is $14/0.467\simeq30 \mathrm{yr}$; the NELIOTA and LRO normalizations are multiplied by $\Gamma(145.5,1)/145$ and $\Gamma(16.5,1)/16$, respectively; and the single-event Juno/UVS normalization is multiplied by $\lambda\sim\Gamma(1.5,1)$.

\begingroup
\newlength{\PriorTableWidth}
\newlength{\PriorRecordWidth}
\newlength{\PriorParameterWidth}
\newlength{\PriorDistributionWidth}
\newlength{\PriorSourceWidth}
\newlength{\PriorGapWidth}
\setlength{\PriorTableWidth}{\textwidth}
\setlength{\PriorRecordWidth}{3.4cm}
\setlength{\PriorParameterWidth}{2.4cm}
\setlength{\PriorSourceWidth}{1.8cm}
\setlength{\PriorGapWidth}{0.18cm}
\setlength{\PriorDistributionWidth}{%
  \dimexpr\PriorTableWidth
  -\PriorRecordWidth
  -\PriorParameterWidth
  -\PriorSourceWidth
  -\PriorGapWidth-\PriorGapWidth-\PriorGapWidth\relax}

\newcommand{\PriorRecordText}[2]{%
  \parbox[t]{\PriorRecordWidth}{%
    \raggedright\strut #1\par\vspace{-0.50ex}%
    {\fontsize{5.9}{6.2}\selectfont\citep{#2}}}}
\newcommand{\PriorRecordHead}[1]{%
  \parbox[t]{\PriorRecordWidth}{\raggedright\strut #1}}
\newcommand{\PriorParameter}[1]{%
  \parbox[t]{\PriorParameterWidth}{\raggedright\strut #1}}
\newcommand{\PriorDistribution}[1]{%
  \parbox[t]{\PriorDistributionWidth}{\raggedright\strut #1}}
\newcommand{\PriorSource}[1]{%
  \parbox[t]{\PriorSourceWidth}{\raggedright\strut #1}}
\newcommand{\PriorCitationLink}[2]{\hyperlink{cite.#1}{#2}}
\newcommand{\PriorLine}[3]{%
  \hbox to \dimexpr\PriorParameterWidth+\PriorGapWidth
                    +\PriorDistributionWidth+\PriorGapWidth
                    +\PriorSourceWidth\relax{%
    \PriorParameter{#1}\hspace{\PriorGapWidth}%
    \PriorDistribution{#2}\hspace{\PriorGapWidth}%
    \PriorSource{#3}\hss}%
  \kern0pt}

\newcommand{\PriorHeaderRow}{%
  \hbox to \PriorTableWidth{%
    \PriorRecordHead{Observation records}\hspace{\PriorGapWidth}%
    \PriorParameter{Parameter}\hspace{\PriorGapWidth}%
    \PriorDistribution{Distribution or adopted treatment}%
    \hspace{\PriorGapWidth}\PriorSource{Source}\hss}}

\newcommand{\PriorRecordRow}[3]{%
  \hbox to \PriorTableWidth{%
    \PriorRecordText{#1}{#2}\hspace{\PriorGapWidth}%
    \vtop{\offinterlineskip #3}\hss}}

\newcommand{\PriorAtomicRecord}[3]{%
  \multicolumn{1}{@{}l@{}}{%
    \vtop{\PriorRecordRow{#1}{#2}{#3}}}\\[2pt]}

\newcommand{\PriorAtomicFirstTargetRecord}[4]{%
  \multicolumn{1}{@{}l@{}}{%
    \vtop{%
      \hbox to \PriorTableWidth{\textit{#1}\hss}\kern1pt
      \PriorRecordRow{#2}{#3}{#4}}}\\[2pt]}

\newcommand{\PriorAtomicTargetRecord}[4]{%
  \multicolumn{1}{@{}l@{}}{%
    \vtop{%
      \hbox to \PriorTableWidth{\rule{\PriorTableWidth}{0.35pt}\hss}\kern1pt
      \hbox to \PriorTableWidth{\textit{#1}\hss}\kern1pt
      \PriorRecordRow{#2}{#3}{#4}}}\\[2pt]}

\startlongtable
\begin{deluxetable*}{l}
\tabletypesize{\scriptsize}
\tablecaption{Prior distributions used in the observational-rate conversion
\label{tab:conversion_priors}}
\tablehead{%
  \multicolumn{1}{@{}l@{}}{\PriorHeaderRow}%
}
\startdata
\PriorAtomicFirstTargetRecord{Earth}{Satellite fireballs}{Brown2002}{%
      \PriorLine{$c_0$}{$\mathcal{N}(1.568,\,0.03)$}{B02}
      \PriorLine{$d_0$}{$\mathcal{N}(2.70,\,0.08)$}{B02}
}
\PriorAtomicRecord{USG decameter bolides}{Chow2025decameter}{%
      \PriorLine{rate at $D>D_{\rm th}$}{$\Gamma(14.5)/(30\ \mathrm{yr})$ (14 events; exposure $14/0.467$)}{C25}
      \PriorLine{$D_{\rm th}$}{$\mathcal{N}(7.5,\,0.5)\,\mathrm{m}$}{C25+A}
      \PriorLine{$b$}{$\mathcal{N}(2.70,\,0.08)$}{B02}
}
\PriorAtomicRecord{Infrasound bolides}{Silber2009}{%
      \PriorLine{amplitude}{2-branch: $4.49\,\exp[\mathcal{N}(0,0.19)]$ (incl. 1963 event)/ $6.50\,\exp[\mathcal{N}(0,0.20)]$ (excl. 1963 event)}{S09+A}
      \PriorLine{$b_E$}{2-branch: $\mathcal{N}(0.603,\,0.055)$}{S09+A}
      \PriorLine{$\rho_i$}{Earth close encounters population mixture (Appendix~\ref{app:density prior})}{PI+A}
      \PriorLine{$v_i$}{Earth mixture (Appendix~\ref{app:velocity priors})}{SIM}
}
\PriorAtomicRecord{USG bolides}{Brown2013}{%
      \PriorLine{$a$}{$\mathcal{N}(3.31,\,0.11)$}{B13}
      \PriorLine{$b_E$}{$\mathcal{N}(0.68,\,0.06)$}{B13}
      \PriorLine{$\rho_i$}{Earth close encounters population mixture (Appendix~\ref{app:density prior})}{PI+A}
      \PriorLine{$v_i$}{Earth mixture (Appendix~\ref{app:velocity priors})}{SIM}
}
\PriorAtomicTargetRecord{Moon}{MSFC flashes}{Suggs2014_LunarFlashFlux}{%
      \PriorLine{$R_0$}{$2.33\times10^4\,\exp[\mathcal{N}(0,0.25)]\,\mathrm{yr^{-1}}$}{S14+A}
      \PriorLine{$M_{\rm thresh}$}{$0.030\,\mathrm{kg}$ (fixed), rescaled by $(\eta_{\rm pub}/\eta)(v_{\rm pub}/v_i)^2$}{S14}
      \PriorLine{$\rho$}{$\rho \sim \mathcal{N}_{>0}(1000, 350)$}{S14+A}
      \PriorLine{$\eta$}{Lognormal with median $1.5\times10^{-3}$ and a factor-of-3 range at $2\sigma$}{S14+A}
      \PriorLine{$b$}{Equal-weight three-component mixture: $\{2.28\pm0.08,\,2.18\pm0.10,\,2.70\pm0.08\}$}{A19\slash S14\slash B02}
      \PriorLine{$v_{\rm pub}$}{$24\,\mathrm{km\,s^{-1}}$ (fixed)}{S14}
      \PriorLine{$v_i$}{Moon close encounters population mixture (Appendix~\ref{app:velocity priors})}{SIM}
}
\PriorAtomicRecord{NELIOTA flashes}{liakos2024neliota}{%
      \PriorLine{count scale}{$\Gamma(145.5)/145$}{L24}
      \PriorLine{$R_0$}{$6.49\times10^4\,\mathrm{yr^{-1}}$ within the reported bin (fixed)}{PII}
      \PriorLine{bin edges}{$0.01$--$0.06\,\mathrm{m}$, rescaled by $(\eta_{\rm pub}/\eta)(v_{\rm pub}/v_i)^2$}{L24}
      \PriorLine{$\rho$}{$\rho \sim \mathcal{N}_{>0}(1800, 630)$}{L24+A}
      \PriorLine{$\eta$}{Lognormal with median $1.5\times10^{-3}$ and a factor-of-3 range at $2\sigma$}{L24+A}
      \PriorLine{$b$}{Same three-component mixture as for MSFC}{A19\slash S14\slash B02}
      \PriorLine{$v_{\rm pub}$}{$17\,\mathrm{km\,s^{-1}}$ (fixed)}{L24}
      \PriorLine{$v_i$}{Moon mixture (Appendix~\ref{app:velocity priors})}{SIM}
}
\PriorAtomicRecord{LRO fresh craters}{Speyerer2016}{%
      \PriorLine{count scale}{$\Gamma(16.5)/16$}{Sp16}
      \PriorLine{$R_0$}{$184\,\mathrm{yr^{-1}}$ at $D_c>10\,\mathrm{m}$ (fixed)}{Sp16}
      \PriorLine{$b$}{$\mathcal{N}(2.28,\,0.08)$}{A19}
      \PriorLine{$D_c/D_i$}{Combined gravity and strength regime $\pi$-scaling with impact angle distributed as $p(\theta)\propto\sin 2\theta$}{HH07\slash SIM}
      \PriorLine{$D_c/D_i$ systematics}{Lognormal $(1,0.20)$}{A}
      \PriorLine{$\rho_i$}{Moon close encounters population mixture (Appendix~\ref{app:density prior})}{PI+A}
      \PriorLine{$v_i$}{Moon mixture (Appendix~\ref{app:velocity priors})}{SIM}
}
\PriorAtomicRecord{Apollo seismic impacts}{lognonneMoonMeteoriticSeismic2009a}{%
      \PriorLine{$F_0$}{$1.29\times10^3\,\exp[\mathcal{N}(0,0.30)]$}{L09+A}
      \PriorLine{$f$}{$\mathcal{N}(0.90,\,0.05)$}{L09}
      \PriorLine{$\rho$}{$\mathcal{N}_{>0}(3000, 600)$}{L09}
}
\PriorAtomicTargetRecord{Mars}{CTX/HiRISE fresh craters}{Daubar2014craters,Daubar2024_MarsSeismicCratering}{%
      \PriorLine{$\Phi$ anchor}{$1.65\times10^{-6}\,\exp[\mathcal{N}(0,0.35)]\,\mathrm{km^{-2}\,yr^{-1}}$}{D24+A}
      \PriorLine{$D_{c,\rm anchor}$}{$3.9\,\mathrm{m}$}{D24}
      \PriorLine{$b_{\rm CTX}$}{$\mathcal{N}(2.5,\,0.2)$}{Z24+A}
      \PriorLine{$D_c/D_i$}{Same $\pi$-scaling treatment as for LRO}{HH07\slash SIM}
      \PriorLine{$D_c/D_i$ systematics}{Lognormal $(1,0.20)$}{A}
      \PriorLine{$\rho_i$}{Mars close encounters population mixture (Appendix~\ref{app:density prior})}{PI+A}
      \PriorLine{$v_i$}{Mars mixture (Appendix~\ref{app:velocity priors})}{SIM}
}
\PriorAtomicRecord{InSight seismic impacts}{zenhausern2024estimate}{%
      \PriorLine{$R_0$}{2-branch: $362\,\exp[\mathcal{N}(0,0.47)]\,\mathrm{yr^{-1}}$ (Seismology); $280\,\exp[\mathcal{N}(0,0.35)]\,\mathrm{yr^{-1}}$ (cratering), at $D_c=8\,\mathrm{m}$}{Z24+A}
      \PriorLine{$b_c$}{2-branch: $\mathcal{N}(2.40,\,0.13)$ (Seismology); $\mathcal{N}(2.55,\,0.10)$ (Cratering)}{Z24}
      \PriorLine{$D_c/D_i$, $\rho_i$, $v_i$, $D_c/D_i$ systematics }{As for CTX}{\nodata}
}
\PriorAtomicTargetRecord{Jupiter}{Ground-based flashes}{2013Hueso,hueso2018small}{%
      \PriorLine{$R_0$}{Log-uniform over $10-65\,\mathrm{yr^{-1}}$ (selection corrected; duty cycle included)}{H18}
      \PriorLine{$D_{0,\rm pub}$}{Log-uniform over $5$--$20\,\mathrm{m}$}{H18}
      \PriorLine{$b$}{$\mathcal{N}(2.1,\,0.2)$}{PII+A}
      \PriorLine{$v_{\rm pub}$}{$60\,\mathrm{km\,s^{-1}}$ (fixed), with correction $(v_{\rm pub}/v_i)^{2/3}$}{H18}
      \PriorLine{$v_i$}{Jupiter mixture (Appendix~\ref{app:velocity priors})}{SIM}
}
\PriorAtomicRecord{Juno/UVS bolide}{giles2021detection}{%
      \PriorLine{$\lambda$}{$\Gamma(1.5)$ (one event)}{G21}
      \PriorLine{$R_0$}{$2.4\times10^4\,\mathrm{yr^{-1}}$ (fixed)}{G21}
      \PriorLine{$m_0$}{Log-uniform over $250$--$5000\,\mathrm{kg}$, with $m\propto v^{-2}$ reassignment}{G21}
      \PriorLine{$\rho$}{Log-uniform over $250$--$2000\,\mathrm{kg\,m^{-3}}$}{G21}
      \PriorLine{$b$}{$\mathcal{N}(2.1,\,0.2)$}{PII+A}
      \PriorLine{$v_{\rm pub}$}{$60\,\mathrm{km\,s^{-1}}$ (fixed)}{G21}
      \PriorLine{$v_i$}{Jupiter mixture (Appendix~\ref{app:velocity priors})}{SIM}
}
\enddata
\tablecomments{\textit{Source codes.}
Citations: B02 = \PriorCitationLink{Brown2002}{Brown et al.\ (2002)};
B13 = \PriorCitationLink{Brown2013}{Brown et al.\ (2013)};
C25 = \PriorCitationLink{Chow2025decameter}{Chow \& Brown (2025)};
S09 = \PriorCitationLink{Silber2009}{Silber et al.\ (2009)};
S14 = \PriorCitationLink{Suggs2014_LunarFlashFlux}{Suggs et al.\ (2014)};
L24 = \PriorCitationLink{liakos2024neliota}{Liakos et al.\ (2024, NELIOTA)};
Sp16 = \PriorCitationLink{Speyerer2016}{Speyerer et al.\ (2016, LRO)};
L09 = \PriorCitationLink{lognonneMoonMeteoriticSeismic2009a}{Lognonn\'e et al.\ (2009, Apollo)};
D24 = \PriorCitationLink{Daubar2014craters}{Daubar et al.\ (2014,}
\PriorCitationLink{Daubar2024_MarsSeismicCratering}{2024, CTX/HiRISE)};
Z24 = \PriorCitationLink{zenhausern2024estimate}{Zenh\"ausern et al.\ (2024, InSight)};
G21 = \PriorCitationLink{giles2021detection}{Giles et al.\ (2021, Juno/UVS)};
H18 = \PriorCitationLink{2013Hueso}{Hueso et al.\ (2013,}
\PriorCitationLink{hueso2018small}{2018)};
A19 = \PriorCitationLink{avdellidouTemperaturesLunarImpact2019}{Avdellidou \& Vaubaillon (2019)};
HH07 = \PriorCitationLink{HolsappleHousen2007_CraterScaling}{Holsapple \& Housen (2007)}.
Other codes:
\textbf{A} = assigned in this work (a documented analysis choice with no
published uncertainty);
\textbf{X+A} = central value from source X with a spread assigned in this work;
\textbf{SIM} = derived from the N-body simulations in Paper~I;
\textbf{PI} = population-mixture weights from Paper~I, Tables~3--4;
\textbf{PII} = value adopted or derived in Paper~II (this work);
\textbf{HH07/SIM} = Holsapple--Housen scaling evaluated with
simulation-derived velocities; and
\textbf{A19/S14/B02} = equal-weight mixture of the three published slopes.}
\end{deluxetable*}
\endgroup

\section{Density Priors}
\label{app:density prior}

We construct density priors from the reported values or density ranges for each observational record, using notation of $\rho$, listed below in the same order as in Appendix~\ref{app:conversion_priors}. For studies without reported density values, we construct the density priors from the population mixture described in Section~\ref{sec:prior}, denoted by $\rho_i$.

\citet{Brown2002} convert bolide energies to diameters assuming a bulk density of $3000~{\rm kg\,m^{-3}}$, chosen as a compromise between ordinary- and carbonaceous-chondrite densities, and we inherit this choice directly and do not include any sampling. \citet{Chow2025decameter} adopt $1500~{\rm kg\,m^{-3}}$, anchored to measured bulk densities of decameter-scale NEAs, and compute each diameter from the reported per-event velocity, and the density is already embedded in the published $7.5~{\rm m}$ threshold. We do not sample it. 
\citet{Suggs2014_LunarFlashFlux} did not assume bulk density in their flux determination, as the $30~{\rm g}$ completeness limit follows directly from kinetic energy at an assumed $24~{\rm km\,s^{-1}}$. However, the impactor diameters in their Table~3 correspond exactly to spheres of $1000~{\rm kg\,m^{-3}}$, and our MSFC density prior (truncated normal, median $1000~{\rm kg\,m^{-3}}$) is anchored to this implicit value. The NELIOTA analyses assume $1800~{\rm kg\,m^{-3}}$ and $17~{\rm km\,s^{-1}}$ for sporadic impactors \citep[following][]{Babadzhanov2009, liakos2024neliota,avdellidouTemperaturesLunarImpact2019}, and we use it in our conversion chain through the $(\rho_{\rm pub}/\rho)^{1/3}$ bin-edge rescaling. For the Apollo seismic row we retain the rocky impactor density of $3000~{\rm kg\,m^{-3}}$ used in the seismic flux calculation of \citet{lognonneMoonMeteoriticSeismic2009a}, and broadened to a truncated normal distribution. \citet{2013Hueso,hueso2018small} quote Jovian impactor diameters of $5$--$20~{\rm m}$ for densities between $250~{\rm kg\,m^{-3}}$ (the SL9-like porous case) and $2000~{\rm kg\,m^{-3}}$, a range consistent with our log-uniform $D_0$ prior as we do not sample density. \citet{giles2021detection} adopt the same $250$--$2000~{\rm kg\,m^{-3}}$ range to convert their $250$--$5000~{\rm kg}$ mass estimate to a $0.6-3.4~{\rm m}$ diameter, which we sample as a log-uniform density prior. 

The remaining records do not report density values or assumptions. The infrasound and USG energy relations
\citep{Silber2009,Brown2013} are published as energy fluxes, and the crater and seismic records \citep{Speyerer2016,Daubar2014craters,
Daubar2024_MarsSeismicCratering,zenhausern2024estimate} report crater diameters. Density assumptions are still required in our conversion---the energy-to-diameter and the $D_c/D_i$ scaling, respectively---and we draw from a per-target source population mixture, weighted by the relative impact counts reported in Paper I. We approximate population-specific bulk-density distributions using normal distributions truncated at zero. For NEOs and MBAs, we adopt taxonomy-based priors from \citet{Carry2012}, who reports $\rho_{\rm C}=1.33\pm0.58~{\rm g\,cm^{-3}}$, and $\rho_{\rm S}=2.72\pm0.54~{\rm g\,cm^{-3}}$ for C- and S-type asteroids. For NEOs drawn from \texttt{NEOMOD3} in Paper I, we use albedo as a simplified compositional proxy, where objects with $p_V<0.1$ are assigned the C-type prior and brighter objects the S-type prior, motivated by the dark and bright albedo components in \texttt{NEOMOD3} \citep{NesvornyNEOMOD3}. For MBAs, we use the taxonomic assignments adopted in Paper~I. For JFCs, we adopt $\rho_{\rm JFC}\sim \mathcal{N}_{>0}(480, 220)~{\rm kg\,m^{-3}}$, following \citet{Groussin2019}. For Centaurs, direct density constraints are sparse and largely limited to large objects, with shape-based estimates of roughly $0.6$--$1.1~{\rm g\,cm^{-3}}$ \citep{FernandezValenzuela2017,Leiva2017,BragaRibas2023}. Rather than introduce an unconstrained size-dependent prior, we adopt the same cometary density distribution as for JFCs, noting that this may underestimate the densities of large ($\gtrsim100~{\rm km}$) Centaurs. Our density priors bracket the assumptions from each observation record, from the cometary $250~{\rm kg\,m^{-3}}$ to the chondritic $3000~{\rm kg\,m^{-3}}$ convention, while tying the central values to the simulated source composition of each target.

\section{Impact-velocity Priors}
\label{app:velocity priors}
\setcounter{equation}{0}

We construct the velocity priors from the Hill-sphere close encounters reweighted by the gravitational-focusing impact probability. Using the specific orbital energy $\epsilon$, speed and periapsis distance from the Paper I simulation, the impact speed for any close encounter, bound or unbound, is
\begin{equation}
v_{\rm imp}=\sqrt{2\epsilon + 2\mu_p/R_p},
\end{equation}
and each close encounter is weighted by impact probability $w\propto b_{\rm crit}^2 = R_p^2\,(1+v_{\rm esc}^2/v_{\rm eff}^2)$, with $v_{\rm eff}=v_\infty=\sqrt{2\epsilon}$ for unbound passages and the planetocentric speed at the Hill radius for bound ones. Repeated close encounters of bound objects appear as repeated rows and thus receive their natural repeat-passage representation.

Per target, populations are mixed by the Paper~I weights of the chosen estimator (periapsis-based: Jupiter $11/1334/1636$ NEO/MBA/JFC, Earth $5/5$ NEO/JFC; B-plane based: Jupiter $7.14/47.9/865/0.22$, Earth $5.50/0.46$; Mars only uses the B-plane weights $0.435/0.029/0.08$ since it has zero periapsis impacts).

\begin{table}[t]
\centering
\caption{Constructed Impact Velocity Prior Distributions}
\label{tab:velocity distribution}
\begin{tabular}{lllll}
\toprule
Target & Basis & $v_{\rm imp}$ median
& 68\% interval
& 95\% interval ($\mathrm{km\,s^{-1}}$) \\
\midrule
Earth   & periapsis & 17.8 & 14.4--24.0 & 12.6--32.8 \\
Earth   & B-plane   & 16.4 & 13.6--22.4 & 12.0--28.7 \\
Moon    & periapsis & 15.3 & 10.3--22.5 & 6.3--33.6 \\
Moon    & B-plane   & 13.2 & 8.8--20.5  & 5.7--28.1 \\
Mars    & B-plane      & 12.8 & 9.1--18.3  & 6.7--24.6 \\
Jupiter & periapsis & 59.5 & 59.5--59.9 & 59.5--60.4 \\
Jupiter & B-plane   & 59.7 & 59.5--60.1 & 59.5--60.6 \\
\bottomrule
\end{tabular}
\end{table}

The resulting distributions can be found in Table~\ref{tab:velocity distribution}. There are three major findings. (i) The Earth median of $16$--$18\,\mathrm{km\,s^{-1}}$ brackets the standard $20\,\mathrm{km\,s^{-1}}$ fireball assumption, and the Earth energy rows are only mildly velocity-sensitive. (ii) The lunar median of $\simeq13$--$15\,\mathrm{km\,s^{-1}}$ is slower than the $24\,\mathrm{km\,s^{-1}}$ (MSFC) and $17\,\mathrm{km\,s^{-1}}$ (NELIOTA) sporadic assumptions. At fixed flash energy, slower projectiles are more massive, which raises the flash-derived rates. (iii) The Jovian distribution is a delta function at $\simeq60\,\mathrm{km\,s^{-1}}$ to within $\pm1\%$ for every source population, which makes the\ $60\,\mathrm{km\,s^{-1}}$ assumption from \citet{giles2021detection} well justified and eliminates velocity as a significant source of uncertainty in the Jovian conversion chain. 





\bibliography{ref}{}
\bibliographystyle{aasjournalv7}



\end{document}